\documentclass[10pt,journal,letterpaper,compsoc]{IEEEtran}

\makeatletter
\def\normalsize{\@setfontsize{\normalsize}{9.5bp}{12.00pt}}
\normalsize \makeatother

\usepackage{graphics}
\usepackage{graphicx,color}
\usepackage{epsfig}
\usepackage{latexsym}
\usepackage{amsmath}
\usepackage{amssymb}
\usepackage{subfigure}
\usepackage{array}
\usepackage{url}

\begin{document}

\title{Characterizing Internet Worm Infection Structure}

\author{Qian~Wang,~\IEEEmembership{Student Member,~IEEE,}
        Zesheng~Chen,~\IEEEmembership{Member,~IEEE,}
        and~Chao~Chen,~\IEEEmembership{Member,~IEEE}

\IEEEcompsocitemizethanks
{\IEEEcompsocthanksitem Q. Wang is
with the Department of Electrical and Computer Engineering,
Florida International University, Miami, FL, 33174.\protect\\
E-mail: qian.wang@fiu.edu. \IEEEcompsocthanksitem Z. Chen and C.
Chen are with the Department of
Engineering, Indiana University - Purdue University Fort Wayne, Fort Wayne, IN 46805.\protect\\
E-mail: \{zchen, chen\}@engr.ipfw.edu. \IEEEcompsocthanksitem Z.
Chen is the Corresponding Author. }}

\bibliographystyle{./IEEEtran}

\IEEEcompsoctitleabstractindextext{
\begin{abstract}
Internet worm infection continues to be one of top security threats
and has been widely used by botnets to recruit new bots. In this
work, we attempt to quantify the infection ability of individual
hosts and reveal the key characteristics of the underlying topology
formed by worm infection, {\em i.e.,} the number of children and the
generation of the worm infection family tree. Specifically, we first
apply probabilistic modeling methods and a sequential growth model to analyze
the infection tree of a wide class of worms. We analytically and
empirically find that the number of children has asymptotically a
geometric distribution with parameter 0.5. As a result, on average
half of infected hosts never compromise any vulnerable host, over
98\% of infected hosts have no more than five children, and a small
portion of infected hosts have a large number of children. We also
discover that the generation follows closely a Poisson distribution
and the average path length of the worm infection family tree
increases approximately logarithmically with the total number of
infected hosts. Next, we empirically study the infection structure
of localized-scanning worms and surprisingly find that most of the
above observations also apply to localized-scanning worms. Finally,
we apply our findings to develop bot detection methods and study
potential countermeasures for a botnet ({\em e.g.,} Conficker C)
that uses scan-based peer discovery to form a P2P-based botnet.
Specifically, we demonstrate that targeted detection that focuses on
the nodes with the largest number of children is an efficient way to
expose bots. For example, our simulation shows that when 3.125\%
nodes are examined, targeted detection can reveal 22.36\% bots.
However, we also point out that future botnets may limit the maximum
number of children to weaken targeted detection, without greatly
slowing down the speed of worm infection.
\end{abstract}

}

\maketitle

\IEEEdisplaynotcompsoctitleabstractindextext
\IEEEpeerreviewmaketitle

\section{Introduction}
\label{sec:intro}

Internet epidemics are malicious software that can self-propagate
across the Internet, {\em i.e.,} compromise vulnerable hosts and use
them to attack other victims. Internet epidemics include viruses,
worms, and bots. The past more than twenty years have witnessed the
evolution of Internet epidemics. Viruses infect machines through
exchanged emails or disks, and dominated the 1980s and 1990s.
Internet active worms compromise vulnerable hosts by automatically
propagating through the Internet and have caused much attention
since the Code Red and Nimda worms in 2001. Botnets are zombie
networks controlled by attackers through Internet relay chat (IRC)
systems ({\em e.g.,} GT Bot) or peer-to-peer (P2P) systems ({\em
e.g.,} Storm) to execute coordinated attacks and have become the
number one threat to the Internet in recent years. Since Internet
epidemics have evolved to become more and more virulent and
stealthy, they have been identified as one of the top security
problems \cite{Grand}.

The main difference between worms and botnets lies in that worms
emphasize the procedures of infecting targets and propagating among
vulnerable hosts, whereas botnets focus on the mechanisms of
organizing the network of compromised computers and setting out
coordinated attacks. Most botnets, however, still apply
worm-scanning methods to recruit new bots or collect network
information \cite{Dagon06,Li09,Vogt_NDSS'07_Army,Wang_HotBots'07_Hybrid}.
Moreover, although many P2P-based botnets use the existing P2P
networks to build a bootstrap procedure, Conficker C forms a P2P
botnet through scan-based peer discovery \cite{SRIcon,CAIDA}.
Specifically, Conficker C searches for new peers by randomly
scanning the entire Internet address space. As a result, the way
that Conficker C constructs a P2P-based botnet is in principle the
same as worm scanning/infection.
Therefore, characterizing the structure of worm infection is important and
imperative for defending against current and future epidemics such
as Internet worms and Conficker C like P2P-based botnets.

\begin{figure}[tb]
\begin{center}
\includegraphics [width= 7.5cm, bb=196 310 486 485]{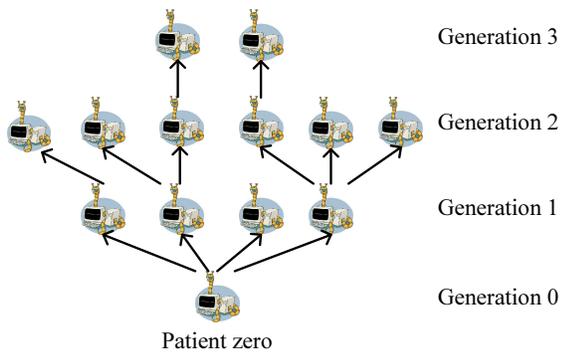}
\caption{A worm tree.} \label{tree}
\end{center}
\end{figure}

Modeling Internet worm infection has been focused on the {\em macro}
level. Most, if not all, mathematical models study the total number
of infected hosts over time
\cite{Staniford,Zou:scan,AAWP,Rohloff,Vojnovic,Dagon06}. The models
of the {\em micro} level of worm infection, however, have been
investigated little. The micro-level models can provide more
insights into the infection ability of individual compromised hosts
and the underlying topologies formed by worm infection. A key
micro-level information is ``who infects whom" or the worm infection
family tree. When a host infects another host, they form a
``father-and-son" relationship, which is represented by a directed
edge in a graph formed by worm infection. Hence, the procedure of
worm propagation constructs a directed tree where patient zero is
the root and the infected hosts that do not compromise any
vulnerable host are leaves (see Fig. \ref{tree}). To the best of our
knowledge, there is yet no mathematical model for reflecting the
structure of such a tree.

The goal of this work is to characterize the Internet worm infection
family tree, {\em i.e.,} the topology formed by worm infection. For
such a tree, we are particularly interested in two metrics:
\begin{itemize}
\item {\it Number of children:}
For a randomly selected node in the tree, how many children does it
have? This metric represents the infection ability of individual
hosts.
\item{\it Generation:}
For a randomly selected node in the tree, which generation (or
level) does it belong to? This metric indicates the average path
length of the graph formed by worm infection.
\end{itemize}
These two metrics reflect the underlying topology formed by worm
infection, called the ``worm tree'' in short. For example, if the
worm tree is a random graph, each host would infect a similar number
of targets; and the average path length would increase approximately
logarithmically with the total number of nodes
\cite{Barabasi_complex,Dagon_ACSAC'07_Taxonomy}. If the worm tree
has a power-law topology, only a very small number of hosts infect a
large number of children, and a majority of hosts infect none or few
children; and the average path length would also increase
approximately logarithmically with the total number of nodes
\cite{Barabasi_complex}. Moreover, power-law topologies are robust
to random node removal, but are vulnerable to the removal of a small
portion of nodes with highest node degrees. However, random graphs
are robust to both removal schemes \cite{Barabasi_complex}.
Therefore, studying the structure of the worm tree can help provide
insights on detecting and defending against botnets such as
Conficker C.

To study these two metrics analytically, we apply probabilistic
modeling methods and derive the joint probability distribution of
the number of children and the generation through a sequential
growth model. Specifically, we start from a worm tree with only
patient zero and add new nodes into the worm tree sequentially. We
then investigate the relationship between the two worm trees before
and after a new node is added. From the joint distribution, we
analyze the marginal distributions of the number of children and the
generation. We also develop closed-form approximations to both
marginal distributions and the joint distribution. Different from
other models that characterize the dynamics of worm propagation
({\em e.g.,} the total number of infected hosts over time), our
sequential growth model aims at capturing the main features of the
topology formed by worm infection ({\em e.g.,} the number of
children and the generation).

As a first attempt, we analyze the worm tree formed by a wide class
of worms such as random-scanning worms \cite{Staniford},
routable-scanning worms \cite{Wu,Zou:scan}, importance-scanning
worms \cite{Chen:IJSN}, OPT-STATIC worms \cite{Milan:ToN}, and
SUBOPT-STATIC worms \cite{Milan:ToN}. For these worms, a new victim
is compromised by each existing infected host with equal
probability. We then verify the analytical results through
simulations. We also employ simulations to investigate worm
infection using localized scanning \cite{Rajab,Chen07}. Finally, we
apply our analysis and observations to develop methods for detecting
bots and study potential countermeasures for a botnet ({\em e.g.,}
Conficker C) that uses scan-based peer discovery to form a P2P-based
botnet.

Through both analytical and empirical study, we make several
contributions from this research as follows. First, if a worm uses a
scanning method for which a new victim is compromised by each
existing infected host with equal probability, the number of
children is shown both analytically and empirically to have
asymptotically a geometric distribution with parameter 0.5. This
means that on average half of infected hosts never compromise any
target and over 98\% of infected hosts have no more than five
children. On the other hand, this also indicates that a small
portion of hosts infect a large number of vulnerable hosts.
Moreover, the generation is demonstrated to closely follow a Poisson
distribution with parameter $H_n-1$, where $n$ is the number of
nodes and $H_n$ is the $n$-th harmonic number \cite{Harmonic}. This
means that the average path length of the worm tree increases
approximately logarithmically with the number of nodes. Second, if a
worm uses localized scanning, the number of children still has
approximately a geometric distribution with parameter 0.5. Moreover,
the generation still follows a Poisson distribution, but with the
parameter depending on the probability of local scanning. Therefore,
most previous observations also apply to localized-scanning worms.
Finally, a direct application of the observations of the worm tree
is on the bot detection in Conficker C like botnets. We show both
analytically and empirically that while randomly examining a small
portion of nodes in a botnet ({\em i.e.,} random detection) can only
expose a limited number of bots, examining the nodes with the
largest number of children ({\em i.e.,} targeted detection) is much
more efficient in detecting bots. For example, our simulation shows
that when 3.125\% nodes are examined, random detection exposes
totally 9.10\% bots, whereas targeted detection reveals 22.36\%
bots. On the other hand, we also point out that future botnets can
potentially use a simple method to weaken the performance of
targeted detection, without greatly slowing down the speed of worm
infection. To the best of our knowledge, this is the first attempt
in understanding and exploiting the topology formed by worm
infection quantitatively.

The remainder of this paper is structured as follows. Section
\ref{sec:back} presents our sequential growth model and assumptions
used in analyzing the worm tree. Section \ref{sec:ana} gives our
analysis on the worm tree. Section \ref{sec:sim} uses simulations to
verify the analytical results and provide observations on the worm
tree using the localized-scanning method. Section \ref{sec:app}
further develops bot detection methods and studies potential
countermeasures by future botnets. Finally, Section
\ref{sec:related} discusses the related work, and Section
\ref{sec:conclusion} concludes this paper.

\section{Worm Tree and Sequential Growth Model}
\label{sec:back}

In this section, we provide the background on the worm tree, and
present the assumptions and the growth model.

An example of a worm tree is given in Fig. \ref{tree}. Here, patient
zero is the root and belongs to generation 0. The tail of an arrow
is from the ``father" or the infector, whereas the head of an arrow
points to the ``son" or the infectee. If a father belongs to
generation $i$, then its children lie in generation $i+1$. In a worm
tree with $n$ nodes, we use $L_n(i,j)$ ($0\le i,j\le n-1$) to denote
the number of nodes that have $i$ children and belong to generation
$j$. Note that $\sum_{i=0}^{n-1}\sum_{j=0}^{n-1}{L_n(i,j)}=n$. We
also use $C_n(i)$ ($i=0, 1, 2, \cdots, n-1$) to denote the number of
nodes that have $i$ children and $G_n(j)$ ($j=0, 1, 2, \cdots, n-1$)
to denote the number of nodes in generation $j$. Moreover,
$L_n(i,j)$, $C_n(i)$, and $G_n(j)$ are random variables. Thus, we
define $p_n(i,j)=\frac{\mbox{E}[L_n(i,j)]}{n}$, representing the
joint distribution of the number of children and the generation.
Similarly, we define $c_n(i)= \frac{\mbox{E}[C_n(i)]}{n}$ to
represent the marginal distribution of the number of children and
$g_n(j)= \frac{\mbox{E}[G_n(j)]}{n}$ to represent the marginal
distribution of the generation. Note that
$c_n(i)=\sum_{j=0}^{n-1}p_n(i,j)$ and
$g_n(j)=\sum_{i=0}^{n-1}p_n(i,j)$.

\begin{figure}[tb]
\begin{center}
    \mbox{
            \subfigure[Extreme case 2.]{\includegraphics[width= 6.5cm, bb=134 380 392 475]{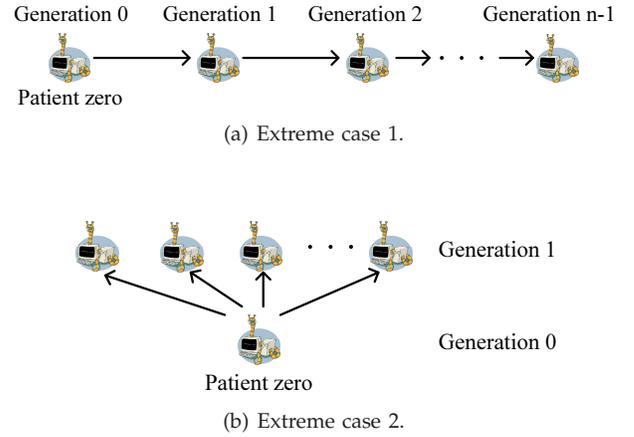}}
      }
          \caption{Two extreme cases of worm trees.}
   \label{fig:extreme}
\end{center}
\end{figure}

Although we model worm infection as a tree, different worm trees can
show very different structures. Fig. \ref{fig:extreme} demonstrates
two extreme cases of worm trees. Specifically, in Fig.
\ref{fig:extreme} (a), each infected host compromises one and only
one host except the last infected host. In this case, if the total
number of nodes is $n$, $C_n(0) = 1$, and $C_n(1) = n-1$, which lead
to $c_n(0)=\frac{1}{n}$ and $c_n(1)=\frac{n-1}{n} \approx 1$ when
$n$ is large. That is, almost each node has one and only one child.
Moreover, $G_n(j)=1$, $j=0,1,2,\cdots,n-1$, which means that
$g_n(j)=\frac{1}{n}$. Thus, the average path length is
$\sum_{j=0}^{n-1}{j\cdot g_n(j)}=\frac{n-1}{2} \sim O(n)$. That is,
the average path length increases linearly with the number of nodes.
Comparatively, Fig. \ref{fig:extreme} (b) shows another case where
all hosts (except patient zero) are infected by patient zero. For
the distribution of the number of children, $c_n(n-1)=\frac{1}{n}$,
and $c_n(0)=\frac{n-1}{n}\approx 1$ when $n$ is large, indicating
that almost every node has no child. For the distribution of the
generation, $g_n(0)=\frac{1}{n}$, and $g_n(1)=\frac{n-1}{n}$, which
leads to that the average path length is $\frac{n-1}{n} \approx 1$
when $n$ is large. Thus, the path length is close to a constant of
1. In this work, we attempt to identify the structure of the worm
tree formed by Internet worm infection.

To study the worm tree analytically, in this paper we make several
assumptions and considerations. First, to simplify the model, we
assume that infected hosts have the same scanning rate. This
assumption is removed in Section \ref{sec:sim}, where we use simulations
to study the effect of the variation of
scanning rates on the worm tree. Second, we consider a wide class of
worms for which a new victim is compromised by each existing
infected host with equal probability. Such worms include
random-scanning worms, routable-scanning worms, importance-scanning
worms, OPT-STATIC worms, and SUBOPT-STATIC worms. Random scanning
selects targets in the IPv4 address space randomly and has been the
main scanning method for both worms and botnets \cite{Staniford,
Li09}; routable scanning finds victims in the routable IPv4 address
space \cite{Wu,Zou:scan}; and importance scanning probes subnets
according to the vulnerable-host distribution \cite{Chen:IJSN}.
OPT-STATIC and SUBOPT-STATIC are optimal and suboptimal scanning
methods that are proposed in \cite{Milan:ToN} to minimize the number
of worm scans required to reach a predetermined fraction of
vulnerable hosts. In Section \ref{sec:ls}, we extend our study to
localized scanning, which preferentially searches for targets in the
local subnet and has also been used by real worms \cite{Rajab,
Chen07}. Third, we consider the classic susceptible $\rightarrow$
infected (SI) model, ignoring the cases that an infected host can be
cleaned and becomes vulnerable again, or can be patched and becomes
invulnerable. The SI model assumes that once infected, a host
remains infected. Such a simple model has been widely applied in
studying worm infection
\cite{Staniford,Zou:scan,Chen07infocom,Milan:ToN}, and presents the
worst case scenario. Fourth, we assume that there is no re-infection. That
is, if an infected host is hit by a worm scan, this host will not be
further re-infected. As a result, every infected host has one and
only one father except for patient zero, and the resulting graph formed by worm infection is a tree.
Fifth, we assume that the worm starts from one infected host, {\em i.e.,} patient zero or a
hitlist size of 1. When the hitlist size is larger than 1, the
underlying infection topology is a worm forest, instead of a worm
tree. Our analysis, however, can easily be extended to model the
worm forest. Finally, to simplify the analysis, we assume that no
two nodes are added to the worm tree at the same time. That is, no
two vulnerable hosts are infected simultaneously. We relax this assumption in Section
\ref{sec:sim} where simulations are performed.
\begin{figure}[tb]
\begin{center}
    \mbox{
       \subfigure[Extreme case 1.]{\includegraphics[width=8cm, bb=46 431 373 491]{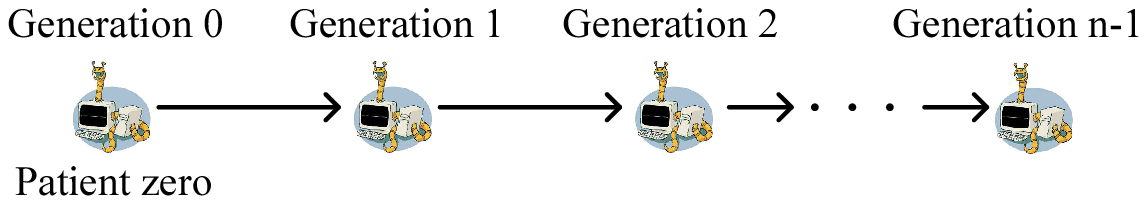}}  }
\end{center}
\end{figure}

Based on these considerations and assumptions, the sequential growth
model of a worm tree works as follows: We consider a fixed sequence
of infected hosts ({\em i.e.,} nodes) $v_1,v_2,\cdots$ and
inductively construct a random worm tree $(T_n)_{n\ge 1}$, where $n$
is the number of nodes and $T_1$ has only patient zero.
Infecting a new host is equivalent to adding a new node into the
existing worm tree. Hence, given $T_{n-1}$, $T_n$ is formed by
adding node $v_n$ together with an edge directed from an existing
node $v_f$ to $v_n$. According to the assumption, $v_f$ is randomly
chosen among the $n-1$ nodes in the tree, {\em i.e.},
$\mbox{Pr}(f=k)=\frac{1}{n-1}$, $k=1,2,\cdots,n-1$. Note that such a
growth model and its variations have been widely used in studying
topology generators \cite{Barabasi_science,Barabasi_Physica_Mean}.
In this paper, we apply this model to characterize worm infection.

\section{Mathematical Analysis}
\label{sec:ana}

In this section, we study the worm tree through mathematical
analysis. Specifically, we first derive the joint distribution of
the number of children and the generation, {\em i.e.,} $p_n(i,j)$,
by applying probabilistic methods. We then use $p_n(i,j)$ to analyze
two marginal distributions, {\em i.e.,} $c_n(i)$ and $g_n(j)$, and
obtain their closed-form approximations. Finally, we find a
closed-form approximation to $p_n(i,j)$.

\subsection{Joint Distribution}

For a worm tree with only patient zero ({\em i.e.,} $n=1$), since
$L_1(0,0)=1$ with probability 1, $p_1(0,0)=1$. Similarly, for a worm
tree with $n=2$, it is evident that $L_2(1,0)=L_2(0,1)=1$. Thus,
$p_2(1,0)=p_2(0,1)=\frac{1}{2}$. We now consider $p_n(i,j)$ ($0\le
i,j \le n-1$) when $n\ge 3$. Specifically, we study two cases:

(1) $p_n(0,j)$, {\em i.e.,} the proportion of the number of leaves
in generation $j$ in $T_n$. Assume that $T_{n-1}$ is given, and
there are $L_{n-1}(0,j)$ leaves in generation $j$ and totally
$G_{n-1}(j-1) = \sum_{i=0}^{n-2}{L_{n-1}(i,j-1)}$ nodes in
generation $j-1$. Note that we have extended the notation so that
$G_{n-1}(-1)=L_{n-1}(i,-1)=0$, $0\le i \le n-2$. When a new node
$v_n$ is added, $v_n$ becomes a leaf of $T_n$. If $v_n$ is connected
to one of existing nodes in generation $j-1$, $v_n$ belongs to
generation $j$; and the probability of such an event is
$\frac{G_{n-1}(j-1)}{n-1}$. Moreover, if a leaf in generation $j$ in
$T_{n-1}$ connects to $v_n$, this node is no longer a leaf and now
has one child; and the probability of this event is
$\frac{L_{n-1}(0,j)}{n-1}$. Therefore, we can obtain the stochastic
recurrence of $L_n(0,j)$:
\begin{equation}
\setlength{\extrarowheight}{0.1cm} L_n(0,j)= \left\{\begin{array} {l
l l}
L_{n-1}(0,j)+1,  & \mbox{w.p.}\ \ \frac{G_{n-1}(j-1)}{n-1} \\
L_{n-1}(0,j)-1,  & \mbox{w.p.}\ \ \frac{L_{n-1}(0,j)}{n-1} \\
L_{n-1}(0,j),    & \ \ \mbox{otherwise}.
\end{array} \right.
\end{equation}
Given $T_{n-1}$ ({\em i.e.,} $L_{n-1}(0,j)$ and $G_{n-1}(j-1)$), the
conditional expected value of $L_n(0,j)$ is
$\left[L_{n-1}(0,j)+1\right]\cdot\frac{G_{n-1}(j-1)}{n-1}+\left[
L_{n-1}(0,j)-1 \right]\cdot \frac{L_{n-1}(0,j)}{n-1} +
L_{n-1}(0,j)\cdot \left[ 1-\frac{G_{n-1}(j-1)+L_{n-1}(0,j)}{n-1}
\right]$, {\em i.e.,}
\begin{equation}
   \textstyle \mbox{E}[L_n(0,j)|T_{n-1}] = \frac{n-2}{n-1}L_{n-1}(0,j)+\frac{1}{n-1}G_{n-1}(j-1).
\end{equation}
Applying $\mbox{E}[L_n(0,j)]=\mbox{E}[\mbox{E}[L_n(0,j)|T_{n-1}]]$
({\em i.e.,} the law of total expectation), we obtain
\begin{equation}
    \textstyle   \mbox{E}[L_n(0,j)] = \frac{n-2}{n-1}\mbox{E}[L_{n-1}(0,j)]+\frac{1}{n-1}\mbox{E}[G_{n-1}(j-1)].
\end{equation}
Using the definitions $p_n(0,j)=\frac{\mbox{E}[L_n(0,j)]}{n}$ and
$g_{n-1}(j-1)=\frac{\mbox{E}[G_{n-1}(j-1)]}{n-1}=\sum_{i=0}^{n-2}{p_{n-1}(i,j-1)}$,
the above equation leads to
\begin{eqnarray}
\label{equ:joint00}
 p_n(0,j) & = &    \textstyle  \frac{n-2}{n}p_{n-1}(0,j)+\frac{1}{n}g_{n-1}(j-1) \\
\label{equ:joint0}
   & = &        \textstyle  \frac{n-2}{n}p_{n-1}(0,j)+\frac{1}{n}\sum_{i=0}^{n-2}{p_{n-1}(i,j-1)}.
\end{eqnarray}

(2) $p_n(i,j)$, $1\le i \le n-1$. Given $L_{n-1}(i,j)$ and
$L_{n-1}(i-1,j)$ in $T_{n-1}$, we study $L_n(i,j)$ in $T_n$. When
the new node $v_n$ is added into $T_{n-1}$, $v_n$ is connected to a
node with $i-1$ children and in generation $j$ with probability
$\frac{L_{n-1}(i-1,j)}{n-1}$, or is connected to a node with $i$
children and in generation $j$ with probability
$\frac{L_{n-1}(i,j)}{n-1}$. Thus, in $T_n$,
\begin{equation}
\setlength{\extrarowheight}{0.1cm} L_n(i,j)= \left\{\begin{array} {l
l l}
L_{n-1}(i,j)+1,  & \mbox{w.p.}\ \ \frac{L_{n-1}(i-1,j)}{n-1} \\
L_{n-1}(i,j)-1,  & \mbox{w.p.}\ \ \frac{L_{n-1}(i,j)}{n-1} \\
L_{n-1}(i,j),    & \ \ \mbox{otherwise}.
\end{array} \right.
\end{equation}
This relationship leads to
\begin{equation}
   \textstyle \mbox{E}[L_n(i,j)|T_{n-1}] = \frac{n-2}{n-1}L_{n-1}(i,j)+\frac{1}{n-1}L_{n-1}(i-1,j).
\end{equation}
Therefore,
\begin{equation}
   \textstyle \mbox{E}[L_n(i,j)] = \frac{n-2}{n-1}\mbox{E}[L_{n-1}(i,j)]+\frac{1}{n-1}\mbox{E}[L_{n-1}(i-1,j)].
\end{equation}
That is,
\begin{equation}
\label{equ:joint1}
   \textstyle  p_n(i,j) = \frac{n-2}{n}p_{n-1}(i,j)+\frac{1}{n}{p_{n-1}(i-1,j)}.
\end{equation}

Summarizing the above two cases, we have the following theorem:
\newtheorem{theorem}{Theorem}
\begin{theorem}
When $n\ge 3$, the joint distribution of the number of children and
the generation in a worm tree $T_n$ follows
\begin{equation}
\setlength{\extrarowheight}{0.1cm} p_n(i,j)= \left\{\begin{array} {l
l l}
\frac{n-2}{n}p_{n-1}(0,j)+\frac{1}{n}g_{n-1}(j-1), & i = 0 \\
\frac{n-2}{n}p_{n-1}(i,j)+\frac{1}{n}{p_{n-1}(i-1,j)}, &
\mbox{otherwise},
\end{array} \right.
\end{equation}
\label{thm:joint}
where $0\le i,j \le n-1$.
\end{theorem}

\begin{figure}[tb]
\begin{center}
{\includegraphics[width=3.5in]{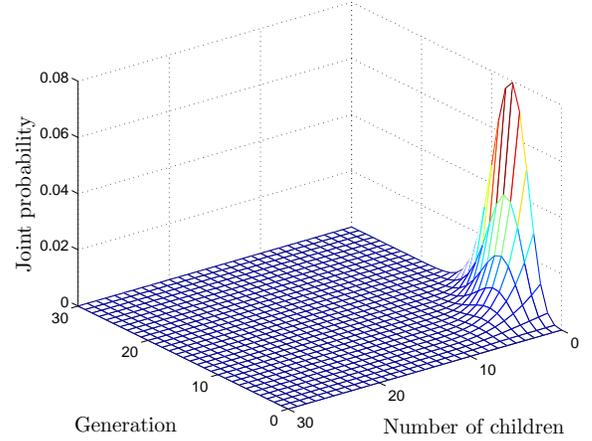}}
          \caption{Joint distribution of the number of children and the generation ($n=2000$).}
\label{fig:joint}
\end{center}
\end{figure}

Theorem \ref{thm:joint} provides a way to calculate $p_n(i,j)$
recursively from $p_2(i,j)$. Fig. \ref{fig:joint} shows a snapshot
of $p_n(i,j)$ when $n=2000$. It can be seen that when the generation
is specified ({\em i.e.,} $j$ is fixed), $p_n(i,j)$ is a monotonous
function and decreases quickly as $i$ increases. On the other hand,
when the number of children is given ({\em i.e.,} $i$ is fixed),
$p_n(i,j)$ has a bell shape. Moreover, since
$\sum_{i=0}^{10}{\sum_{j=0}^{15}{p_n(i,j)}}=0.9976$, most nodes do
not have a large number of children, and the worm tree does not have
a large average path length.

\subsection{Number of Children}

We use $p_n(i,j)$ to derive the marginal distribution of the number
of children, {\em i.e.,} $c_n(i)$. Similarly, we study two cases:

(1) $c_n(0)$, {\em i.e.,} the proportion of the number of leaves in
$T_n$. Since $c_n(0)=\sum_{j=0}^{n-1}{p_n(0,j)}$ and
$\sum_{j=0}^{n-1}g_{n-1}(j-1)=1$, we obtain the recursive
relationship of $c_n(0)$ from Equation (\ref{equ:joint00}):
\begin{equation}
   \textstyle  c_n(0) = \frac{n-2}{n}c_{n-1}(0)+\frac{1}{n}.
\end{equation}
Moreover, note that $c_2(0)=\frac{1}{2}$. If we assume that
$c_{n-1}(0)=\frac{1}{2}$, we can obtain by induction that
\begin{equation}
    c_n(0) =   \textstyle   \frac{1}{2}.
\end{equation}
This indicates that no matter how many nodes are in the worm tree,
on average half of nodes are leaves, {\em i.e.,} on average 50\% of
infected hosts never compromise any target.

(2) $c_n(i)$, $1\le i \le n-1$. From Equation (\ref{equ:joint1}) and
$c_n(i)=\sum_{j=0}^{n-1}{p_n(i,j)}$, we find the recurrence of
$c_n(i)$ as follows
\begin{equation}
\label{equ:rc}
      \textstyle   c_n(i) = \frac{n-2}{n}c_{n-1}(i)+\frac{1}{n}c_{n-1}(i-1).
\end{equation}

Summarizing the above two cases, we have the following theorem on the
distribution of the number of children:
\begin{theorem}
When $n\ge 3$, the distribution of the number of children in a worm
tree $T_n$ follows
\begin{equation}
\setlength{\extrarowheight}{0.1cm} c_n(i)= \left\{\begin{array} {l l
l}
\frac{1}{2},  & i = 0 \\
\frac{n-2}{n}c_{n-1}(i)+\frac{1}{n}c_{n-1}(i-1), & 1 \le i \le n-1.
\end{array} \right.
\end{equation}
\label{thm:pc}
\end{theorem}

From Theorem \ref{thm:pc}, we can derive the statistical properties
of the number of children as follows.
\newtheorem{corollary}{Corollary}
\begin{corollary}
When $n\ge 1$, the expectation and the variance of the number of
children are
\begin{equation}
       \textstyle  \mbox{E}_n[C] = \sum_{i=0}^{n-1}{i\cdot c_n(i)} = \frac{n-1}{n}
\end{equation}
\begin{equation}
     \textstyle  \mbox{Var}_n[C] = \sum_{i=0}^{n-1}{\left(i-\mbox{E}_n[C]
\right)^2 \cdot  c_n(i)} = 2-\frac{n-1}{n^2} - \frac{2H_n}{n},
\end{equation}
where $H_n=\sum\nolimits_{i = 1}^{n} {\frac{1}{i}}$ is the $n$-th
harmonic number \cite{Harmonic}.
\label{cor:C}
\end{corollary}
\vspace{0.1cm}

The proof of Corollary \ref{cor:C} is given in Appendix 1. One
intuitive way to derive $\mbox{E}_n[C]$ is that in worm tree $T_n$,
there are $n-1$ directed edges and $n$ nodes. Thus, the average
number of edges ({\em i.e,} the average number of children) of a
node is $\frac{n-1}{n}$. Moreover, since $H_n$ is $O(1+\ln{n})$,
$\mathop {\lim }\limits_{n \to \infty } \mbox{E}_n[C] = 1$, and
$\mathop {\lim }\limits_{n \to \infty } \mbox{Var}_n[C] = 2$.

Theorem \ref{thm:pc} also leads to a simple closed-form expression
of the distribution of the number of children when $n$ is very
large, as shown in the following corollary.
\begin{corollary} When $n \rightarrow \infty$,
the number of children has a geometric distribution with parameter
$\frac{1}{2}$, {\em i.e.},
    \begin{equation}
    \label{equ:geo}
 c(i) = \mathop {\lim }\limits_{n \to \infty } c_n(i) =  \Big(\frac{1}{2}\Big)^{i+1}, \ \  i = 0, 1, 2, \cdots.
     \end{equation}
     \label{cor:geo}
\end{corollary}

The proof of Corollary \ref{cor:geo} is given in Appendix 2.
Corollary \ref{cor:geo} indicates that when $n$ is very large,
$c_n(i)$ decreases approximately exponentially with a decay constant
of $\ln 2$ as the number of children increases.

\begin{figure}[tb]
\begin{center}
{\includegraphics[width=3.3in]{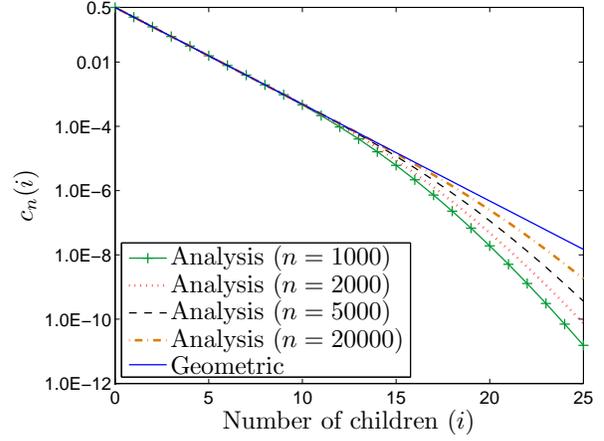}} \caption{Distribution of
the number of children.} \label{fig:children}
\end{center}
\end{figure}

We further study when both $n$ and $i$ are finite and large, how
$c_n(i)$ varies with $n$, {\em i.e.,} how the tail of the
distribution of the number of children changes with $n$. First, note
that $c_3(0)=\frac{1}{2}$, $c_3(1)=\frac{1}{3}$, and
$c_3(2)=\frac{1}{6}$. Thus, from Equation (\ref{equ:rc}), we can
prove by induction that $c_n(i)$ ($n\ge 3$) is a decreasing function
of $i$, {\em i.e.,} $c_n(i) < c_n(i-1)$, for $1\le i \le n-1$. Next,
putting this inequality into Equation (\ref{equ:rc}), we have
$c_n(i)>\frac{n-1}{n}c_{n-1}(i)$. Hence, when $n$ is very large,
$\frac{n-1}{n}\approx 1$, and $c_n(i)>c_{n-1}(i)$, which indicates
that the tail of $c_n(i)$ increases with $n$. Fig.
\ref{fig:children} verifies this result, showing $c_n(i)$ obtained
from Theorem \ref{thm:pc} when $n=1000$, $2000$, $5000$, and
$20000$, as well as the geometric distribution with parameter 0.5
obtained from Corollary \ref{cor:geo}. Note that the y-axis uses
log-scale. It can be seen that when $n$ increases from 1000 to
20000, the tail of $c_n(i)$ also increases to approach the tail of
the geometric distribution. Moreover, it is shown that the geometric
distribution well approximates the distribution of the number of
children when $n$ is large.

\subsection{Generation}

Next, we derive the generation distribution ({\em i.e.,} $g_n(j)$)
in a similar manner to the case of $c_n(i)$. Using Theorem
\ref{thm:joint} and $g_n(j)= \sum_{i=0}^{n-1}{p_n(i,j)}$, we obtain
the following theorem:
\begin{theorem}
When $n\ge 3$, the distribution of the generation in a worm tree
$T_n$ follows
\begin{equation}
\label{equ:rg}
      \textstyle   g_n(j) =
      \frac{n-1}{n}g_{n-1}(j)+\frac{1}{n}g_{n-1}(j-1), 0 \le j \le
      n-1,
\end{equation}
where $g_{n-1}(-1)=0$.
\label{thm:pg}
\end{theorem}

Theorem \ref{thm:pg} gives a method to calculate the distribution of
the generation recursively. Moreover, from Theorem \ref{thm:pg}, we
can derive the statistical properties of the generation distribution
in the following corollary.
\begin{corollary}
When $n\ge 1$, the expectation and the variance of the generation are
\begin{equation}
       \textstyle \mbox{E}_n[G] =  \sum_{j=0}^{n-1}{j\cdot g_n(j)} = H_{n}-1.
\end{equation}
\begin{equation}
     \textstyle \mbox{Var}_n[G] = \sum_{j=0}^{n-1}{\left(j-\mbox{E}_n[G] \right)^2 \cdot g_n(j)} = H_{n}-H_{n,2},
\end{equation}
where $H_n=\sum\nolimits_{i = 1}^{n} {\frac{1}{i}}$ and $H_{n,2} =
\sum\nolimits_{i = 1}^{n} {\frac{1}{i^2}}$. \label{cor:G}
\end{corollary}

The proof of Corollary \ref{cor:G} is given in Appendix 3. From
Corollary \ref{cor:G}, we have some interesting observations. Since
$H_{n}$ is $O(1+ \ln{n})$ and $ H_{\infty,2} = \zeta(2) =
\frac{\pi^2}{6} \approx 1.645$ is the Riemann zeta function of 2
\cite{Riemann1}, both $\mbox{E}_n[G]$ and $\mbox{Var}_n[G]$ are
$O(1+\ln{n})$. This indicates that the average path length of the
worm tree ({\em i.e.}, $\mbox{E}_n[G]$) increases approximately
logarithmically with $n$. Moreover, when $n \rightarrow \infty$,
$\mathop {\lim }\limits_{n \to \infty } \mbox{E}_n[G] - \ln{n} =
\gamma-1$, and $\mathop {\lim }\limits_{n \to \infty }
\mbox{Var}_n[G] -\ln{n} = \gamma - \zeta(2)$, where $\gamma \approx
0.577$ is the Euler-Mascheroni constant \cite{Euler1}. Therefore,
when $n$ is large, $\mbox{E}_n[G] \approx \mbox{Var}_n[G]$.
Furthermore, we can use Theorem \ref{thm:pg} to obtain a closed-form
approximation to $g_n(j)$ as follows.
\begin{corollary}
\label{cor:poi} When $n$ is very large, the generation distribution
$g_n(j)$ can be approximated by a Poisson distribution with
parameter $\lambda_n = \mbox{E}_n[G] = H_n-1$. That is,
\begin{equation}
\label{equ:rg_P}
  \textstyle   g_n(j) \approx \frac{\lambda_n^j}{j!}e^{-\lambda_n}, \ \ 0 \le j \le n-1.
\end{equation}
\label{cor:G2}
\end{corollary}

\begin{figure}[tb]
\begin{center}
{\includegraphics[width=3.3in]{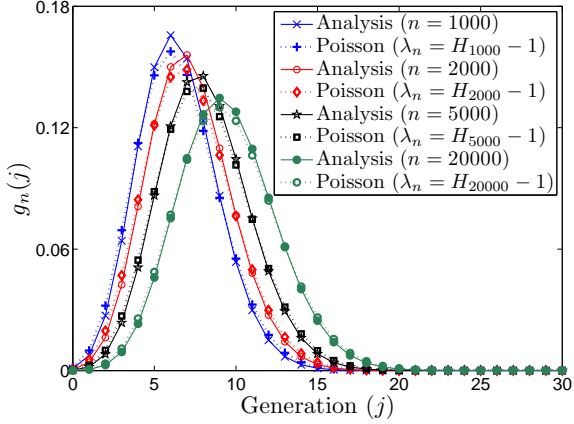}}
          \caption{Distribution of the generation.}
          \label{fig:generation}
\end{center}
\end{figure}

The proof of Corollary \ref{cor:poi} is given in Appendix 4. Fig.
\ref{fig:generation} verifies Corollary \ref{cor:poi}, showing
$g_n(j)$ obtained from Theorem \ref{thm:pg} when $n=1000$, $2000$,
$5000$, and $20000$, as well as the Poisson distribution with
parameter $\mbox{E}_n[G]$. It can be seen that when $n$ is large,
the Poisson distribution fits the generation distribution closely.

\subsection{Approximation to the Joint Distribution}
\begin{figure}[tb]
\begin{center}
{\includegraphics[width=2.8in]{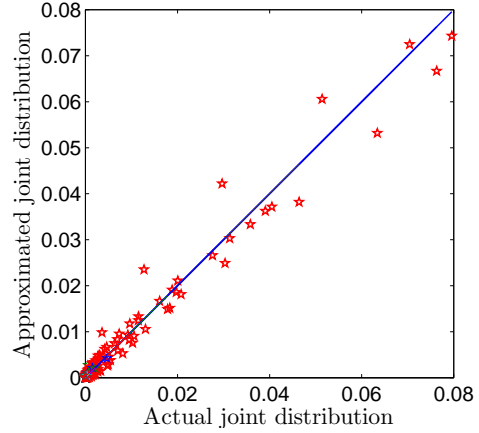}}
          \caption{Parity plot of the approximation to the joint distribution ($n=2000$).}
          \label{fig:approximation}
\end{center}
\end{figure}

Finally, we derive a closed-form approximation to the joint
distribution $p_n(i,j)$. From Equation (\ref{equ:joint1}), we can
see that when $n \to \infty $, $p_n(i,j)=p_{n-1}(i,j)$, which yields
\begin{equation}
   \textstyle  p_n(i,j) = \frac{1}{2}p_n(i-1,j).
\end{equation}
Hence, we can obtain
\begin{equation}
   \textstyle  p_n(i,j) = \left(\frac{1}{2}\right)^i p_n(0,j) \approx \left(\frac{1}{2}\right)^{i+1} g_n(j).
\end{equation}
Since when $n$ is very large, $g_n(j)$ follows closely the Poisson
distribution as in Corollary \ref{cor:poi},
\begin{equation}
\label{equ:joint_app}
   \textstyle  p_n(i,j) \approx \left(\frac{1}{2}\right)^{i+1} \cdot
   \frac{\lambda_n^j}{j!}e^{-\lambda_n}, \ \ 0 \le i,j \le n-1,
\end{equation}
where $\lambda_n = H_n-1$. The above derivation also shows that when
$n$ is very large, the number of children and the generation are
almost independent random variables.

Fig. \ref{fig:approximation} shows the parity plot of the
approximation to the joint distribution when $n=2000$. In the
figure, the x-axis is the actual $p_n(i,j)$ obtained from Theorem
\ref{thm:joint}, and the y-axis is the approximated $p_n(i,j)$ from
Equation (\ref{equ:joint_app}), where $0\le i, j \le 30$. It can be
seen that most points are on or near the diagonal line, indicating
that the approximation to the joint distribution is reasonable.

\begin{figure*}[tb]
\begin{center}
    \mbox{
         \hspace{-0.5cm}
      \subfigure[Number of children.]{\includegraphics[width=2.7in]{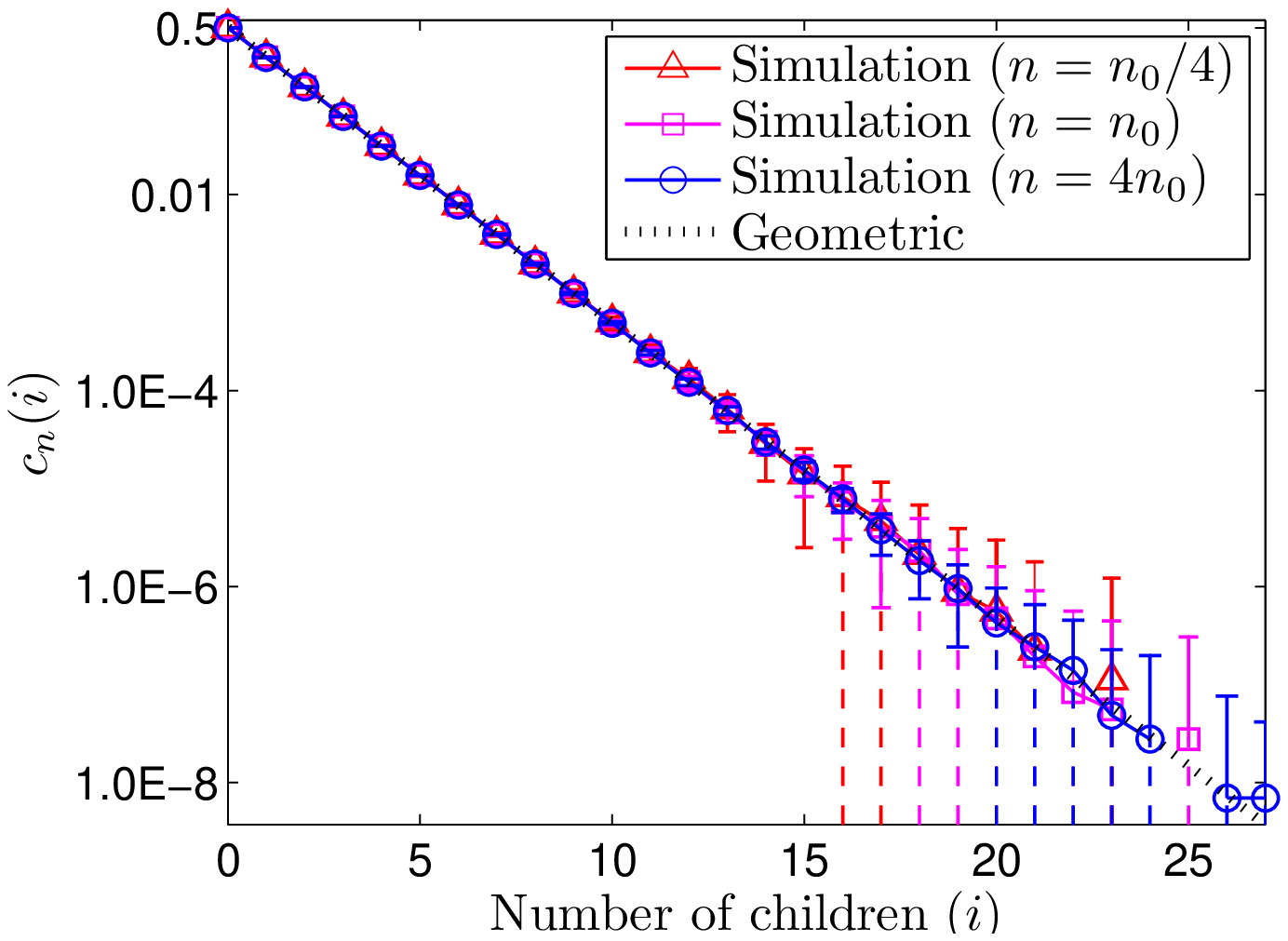}}
      \hspace{-0.7cm}
      \subfigure[Generation.]{\includegraphics[width=2.83in]{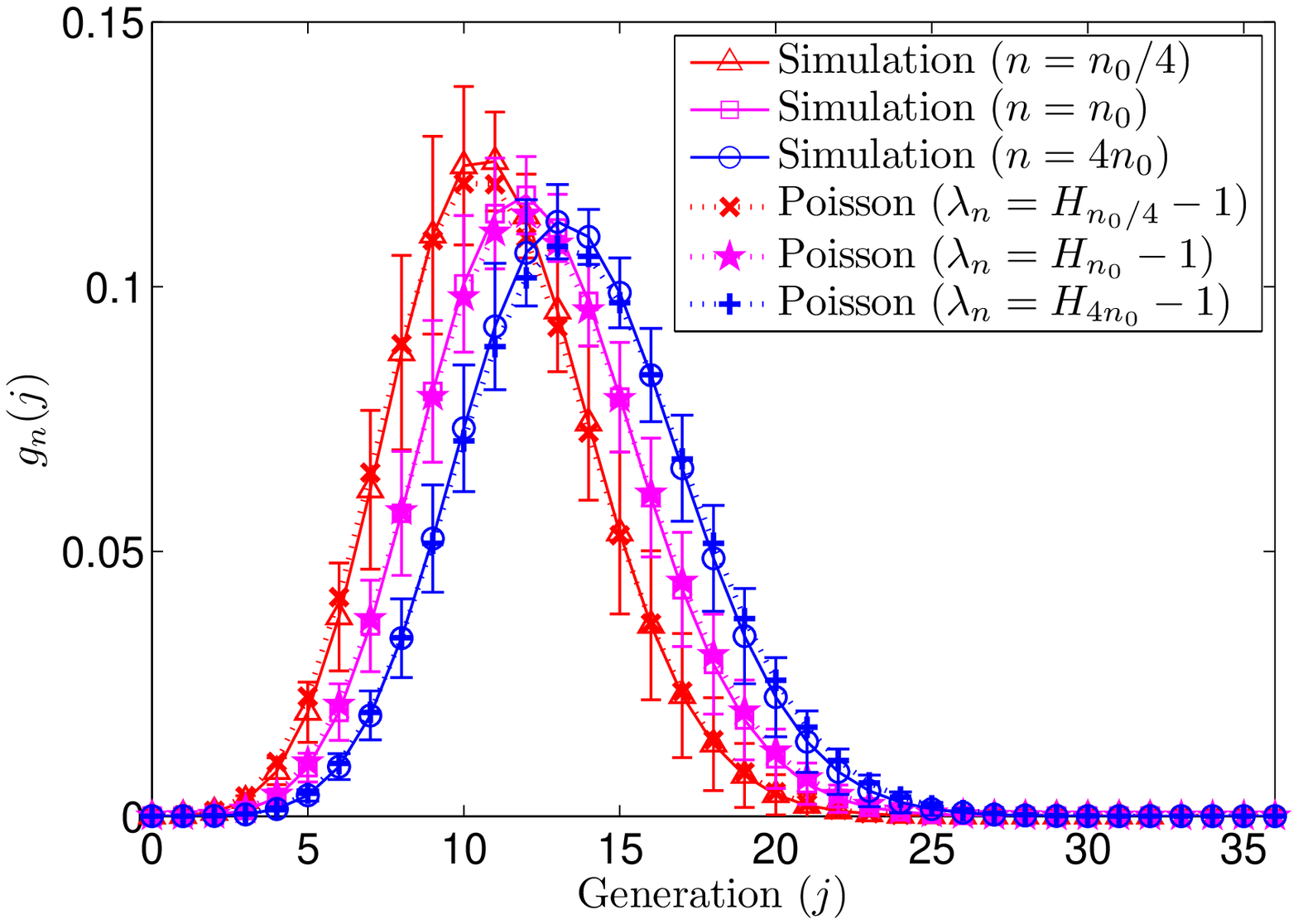}}
      \hspace{-0.7cm}
       \subfigure[Joint distribution ($n=n_0$).]{\includegraphics[width=2.55in]{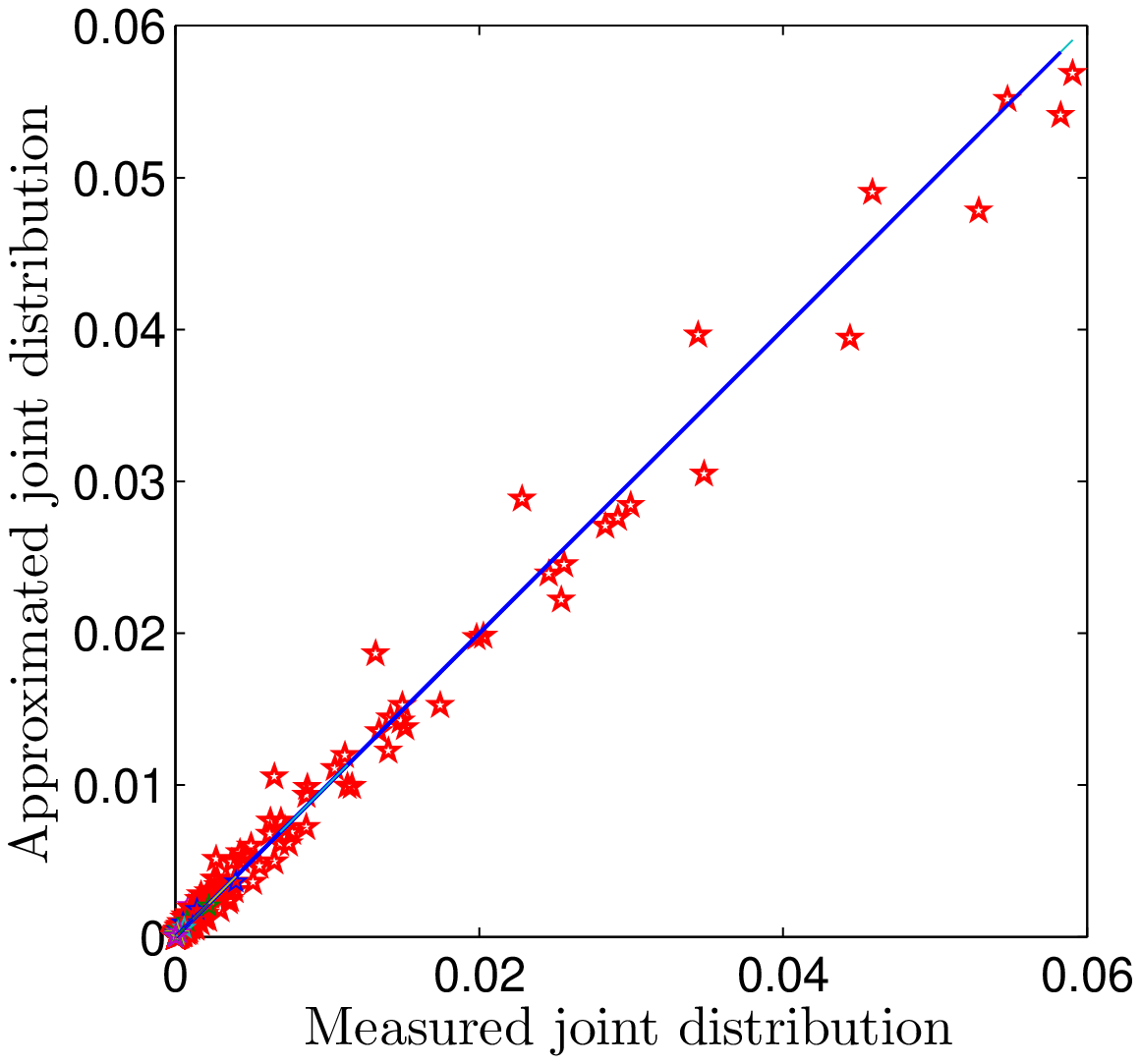}}
      }
          \caption{Simulating the infection structure of the Code Red v2 worm ($n_0=360000$).}
   \label{fig:simu_rs}
\end{center}
\end{figure*}

\section{Simulations and Verification}
\label{sec:sim}

In this section, we study the worm infection structure through
simulations. As far as we know, there is no publicly available data
to show the real worm tree and verify our analytical results.
Moreover, real experiments in a controlled environment are
impractical for this study since the closed-form approximations are
derived based on the assumption that the number of nodes is very
large. Therefore, we apply empirical simulations. Specifically, we
first simulate the infection structure of the Code Red v2 worm and
then study the effects of important parameters on the worm tree.
Finally, we extend our simulation to localized-scanning worms.

\subsection{Code Red v2 Worm Verification}
\label{sec:sim_ver}

We simulate the propagation of the Code Red v2 worm by using and
extending the simulator in \cite{Simulator}. Specifically, the
simulator considers a discrete-time system and mimics the
random-scanning behavior of infected hosts during each discrete time
interval. Moreover, the parameter setting is based on the Code Red
v2 worm's characteristics. For example, the vulnerable population is
$n_0=360,000$, and a newly infected host is assigned with a scanning
rate of 358 scans/min. Detailed information about how the parameters
are chosen can be found in Section VII of \cite{Zou05}. We then
extend the simulator to track the worm infection structure by adding
the information of the number of children and the generation to each
infected host. Moreover, we set the time unit to 20 seconds and
start our simulation at time tick 0 with patient zero. Note that we
remove the assumption used in the sequential growth model that no
two hosts are compromised at the same time. That is, multiple
hosts can be compromised at one time tick. Moreover, all new victims
of the current time tick start scanning at the next time tick. The
simulation results (mean $\pm$ standard deviation) are obtained from
100 independent runs with different seeds and are presented in Fig.
\ref{fig:simu_rs}.

Fig. \ref{fig:simu_rs}(a) shows the distribution of the number of
children, comparing the simulation results of $c_n(i)$ for
$n={n_0}/{4}$, $n_0$, and $4n_0$ with the geometric distribution
obtained from Corollary \ref{cor:geo}. Note that the y-axis uses the
log-scale. The dotted line represents the standard deviation that
goes into the negative territory. It can be seen that the
distribution of the number of children can be well approximated by
the geometric distribution with parameter 0.5. This implies that
$c_n(i)$ decreases approximately exponentially with a decay constant
of $\ln{2}$. Specifically, in all three cases, on average 50.0\% of
the infected hosts do not have children, about 98.4\% of them have
no more than five children, and 0.1\% of them have no less than ten
children. We also calculate the expectation and the variance of the
number of children from the simulation and find that they are
identical to the analytical results obtained from Corollary
\ref{cor:C}. Fig. \ref{fig:simu_rs}(b) demonstrates the generation
distribution, comparing the simulation results of $g_n(j)$ for
$n={n_0}/{4}$, $n_0$, and $4n_0$ with the Poisson distributions with
parameter $\mbox{E}_n[G]=H_n -1$ obtained from Corollary
\ref{cor:poi}. It can be seen that the simulation results of
$g_n(j)$ closely follow the Poisson distributions for all three
cases. Hence, simulation results verify that the average path length
of the worm tree increases approximately logarithmically with the
total number of infected hosts. Moreover, we also compute the
expectation and the variance of the generation in simulations and
verify the analytical results in Corollary \ref{cor:G}. Fig.
\ref{fig:simu_rs}(c) compares the measured joint distribution from
simulations with the approximated joint distribution from Equation
(\ref{equ:joint_app}) by using the parity plot. It can be seen that
most points are on or near the diagonal line, indicating that the
approximation works well.

\subsection{Effects of Worm Parameters}

\begin{figure*}[tb]
\begin{center}
    \mbox{
          \hspace{-0.9cm}
       \subfigure[Effect of scanning rates on $c_n(i)$.]{\includegraphics[width=2.6in]{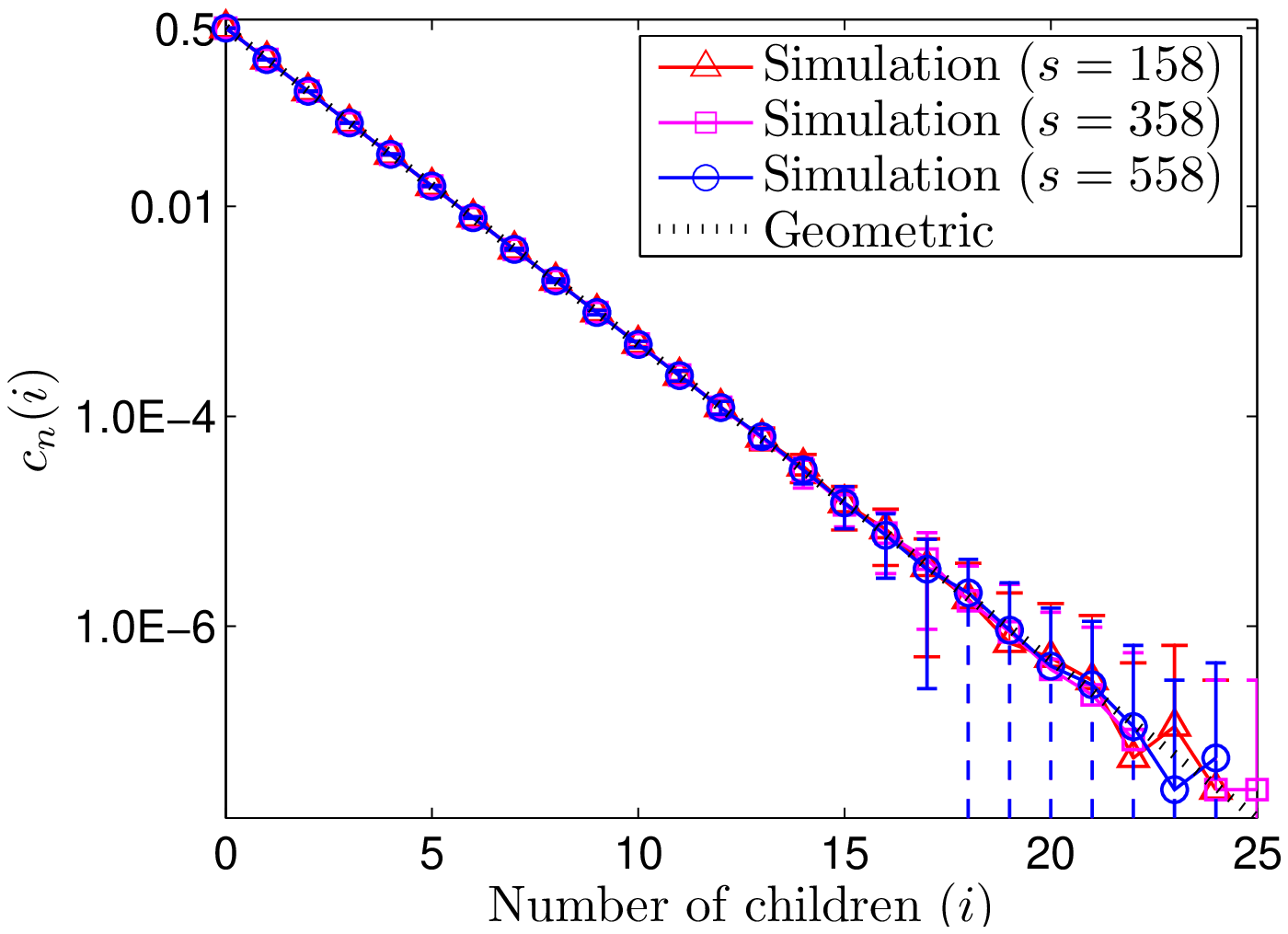}}
            \hspace{-0.7cm}
        \subfigure[Effect of scanning rates on $g_n(j)$.]{\includegraphics[width=2.73in]{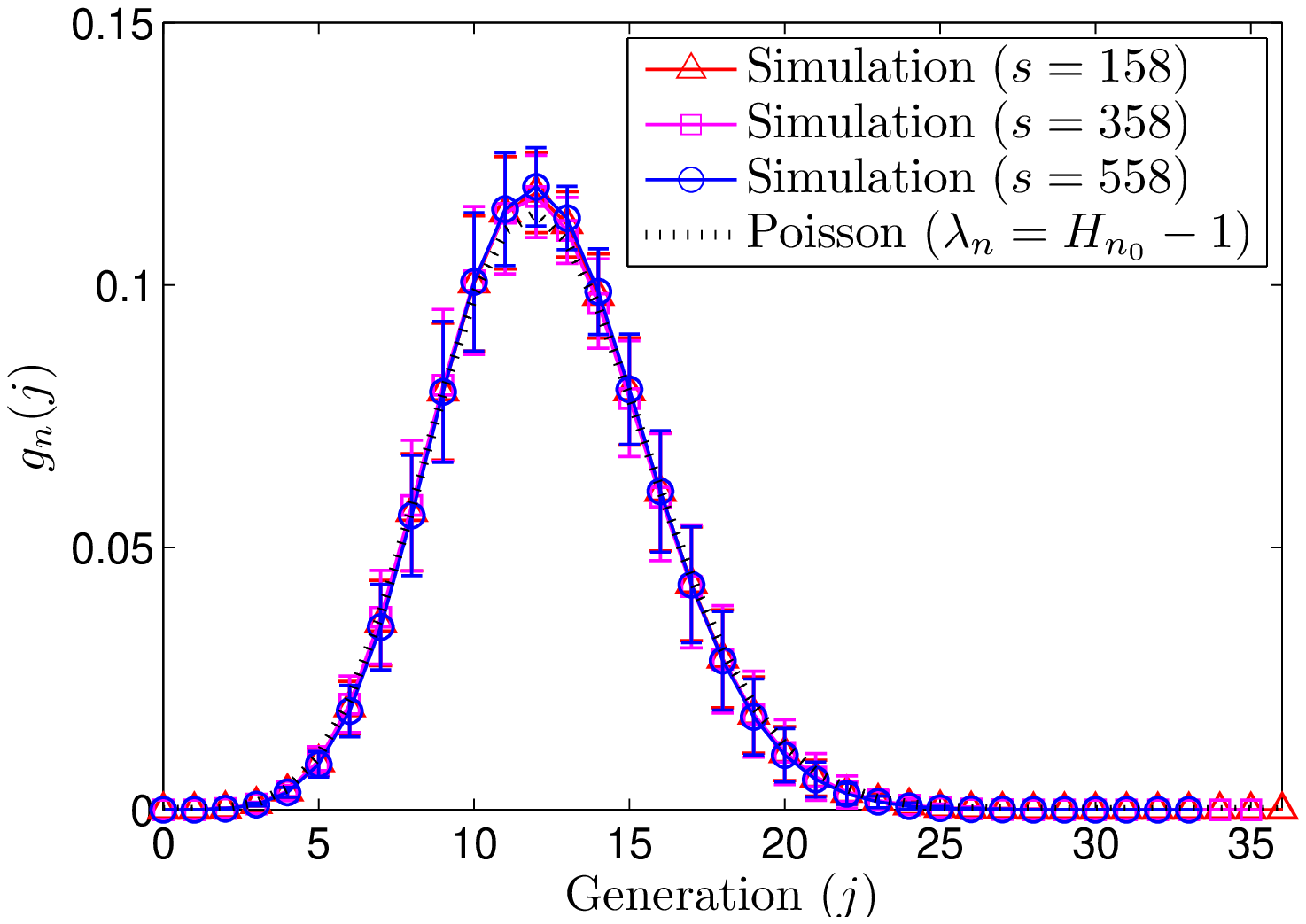}}
                \hspace{-0.7cm}
       \subfigure[Effect of scanning rate standard deviation on $c_n(i)$.]{\includegraphics[width=2.6in]{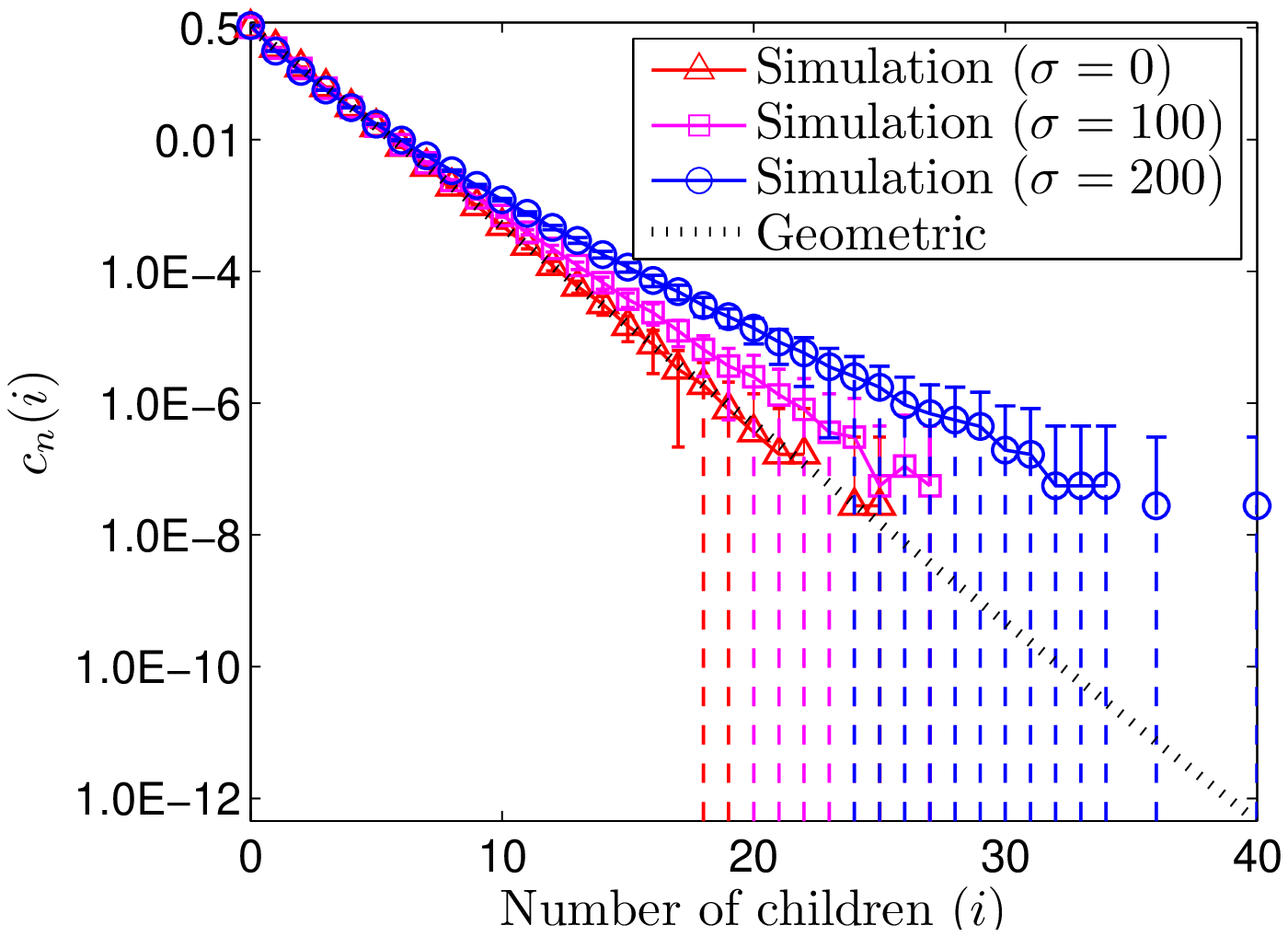}}
      }
\end{center}
\end{figure*}
\begin{figure*}[tb]
\begin{center}
    \mbox{
          \hspace{-0.6cm}
       \subfigure[Effect of scanning rate standard deviation on $g_n(j)$.]{\includegraphics[width=2.6in]{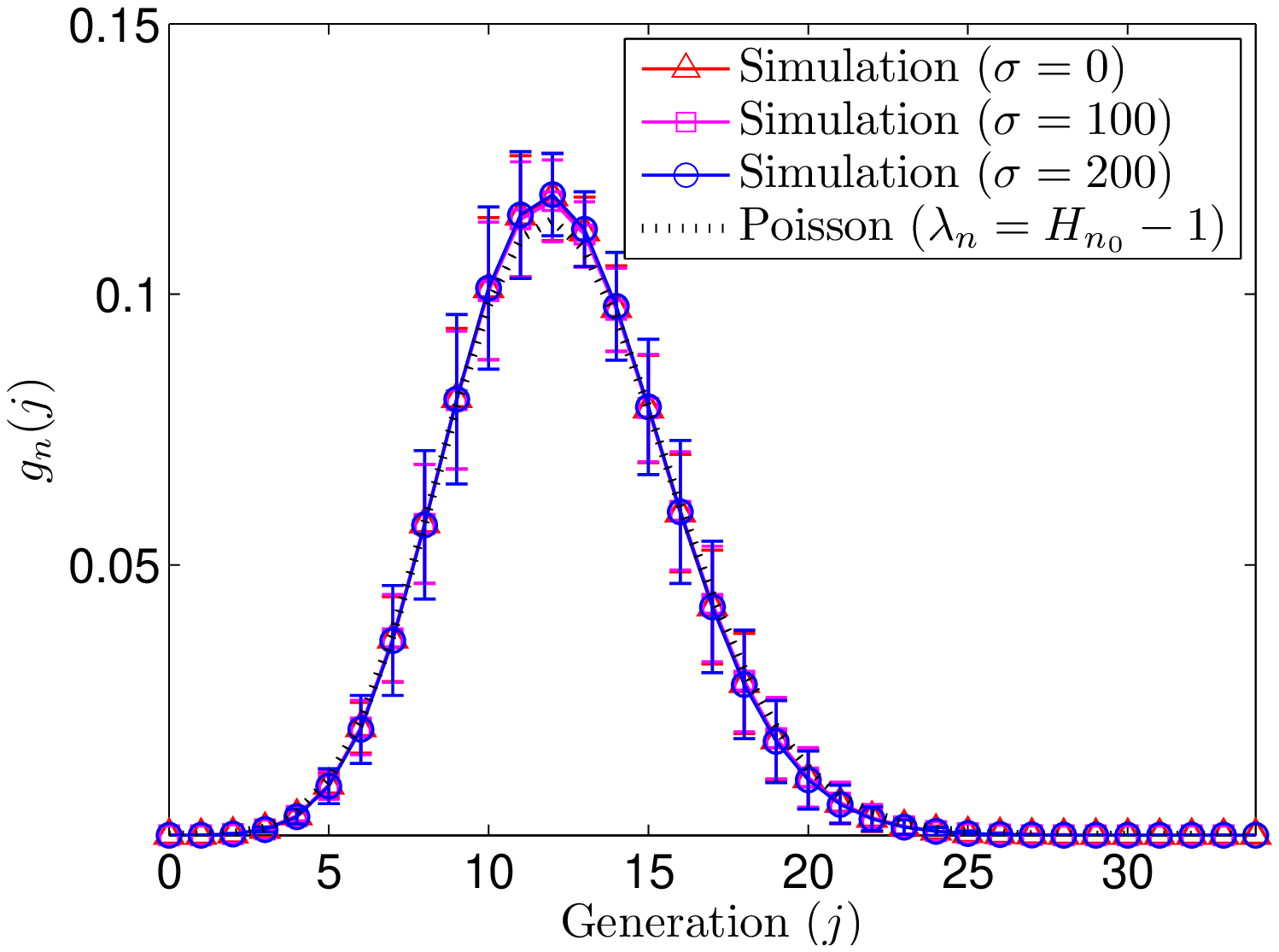}}
            \hspace{-0.7cm}
        \subfigure[Effect of hitlist sizes on $c_n(i)$.]{\includegraphics[width=2.6in]{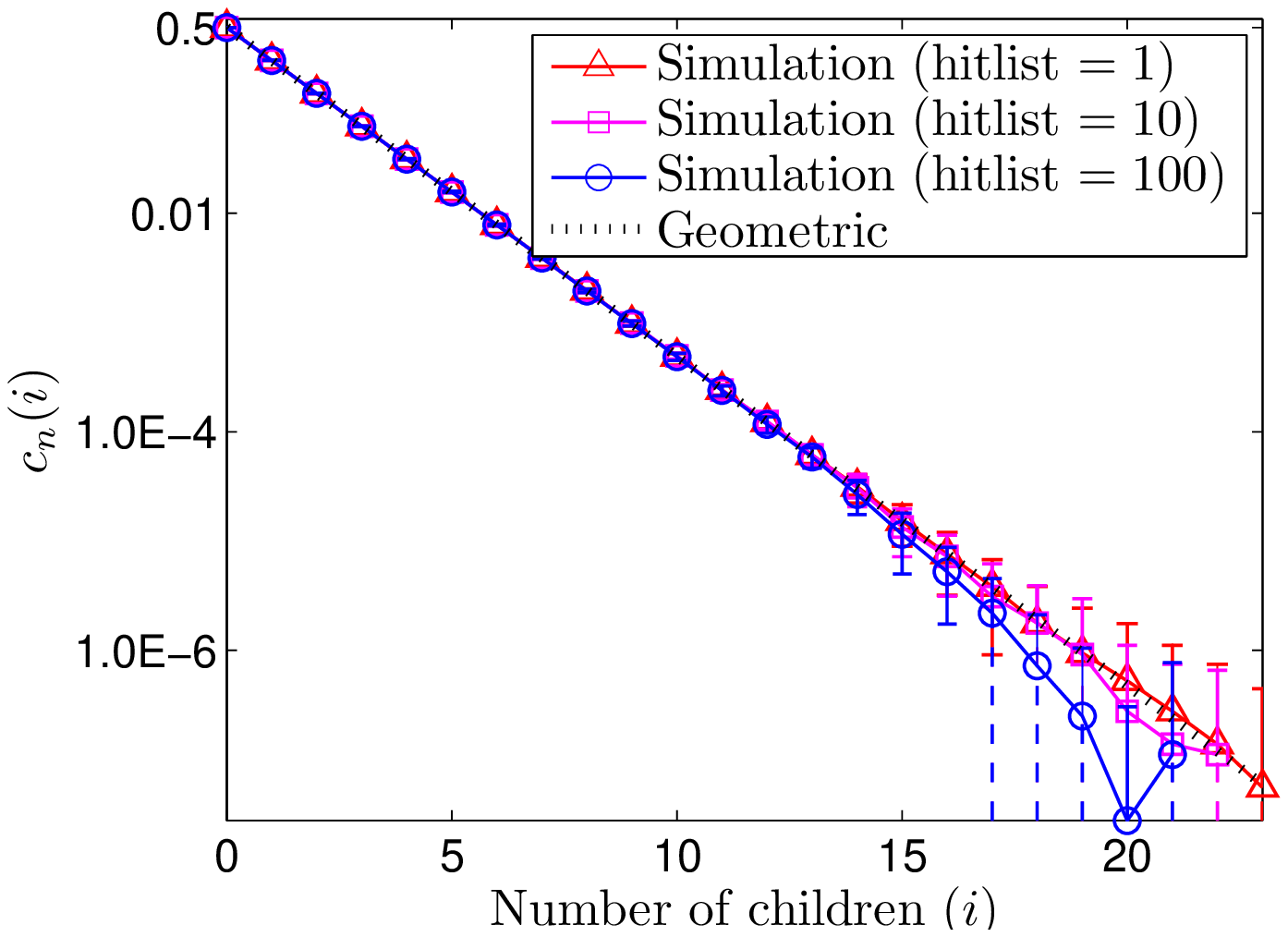}}
                \hspace{-0.85cm}
       \subfigure[Effect of hitlist sizes on $g_n(j)$.]{\includegraphics[width=2.63in]{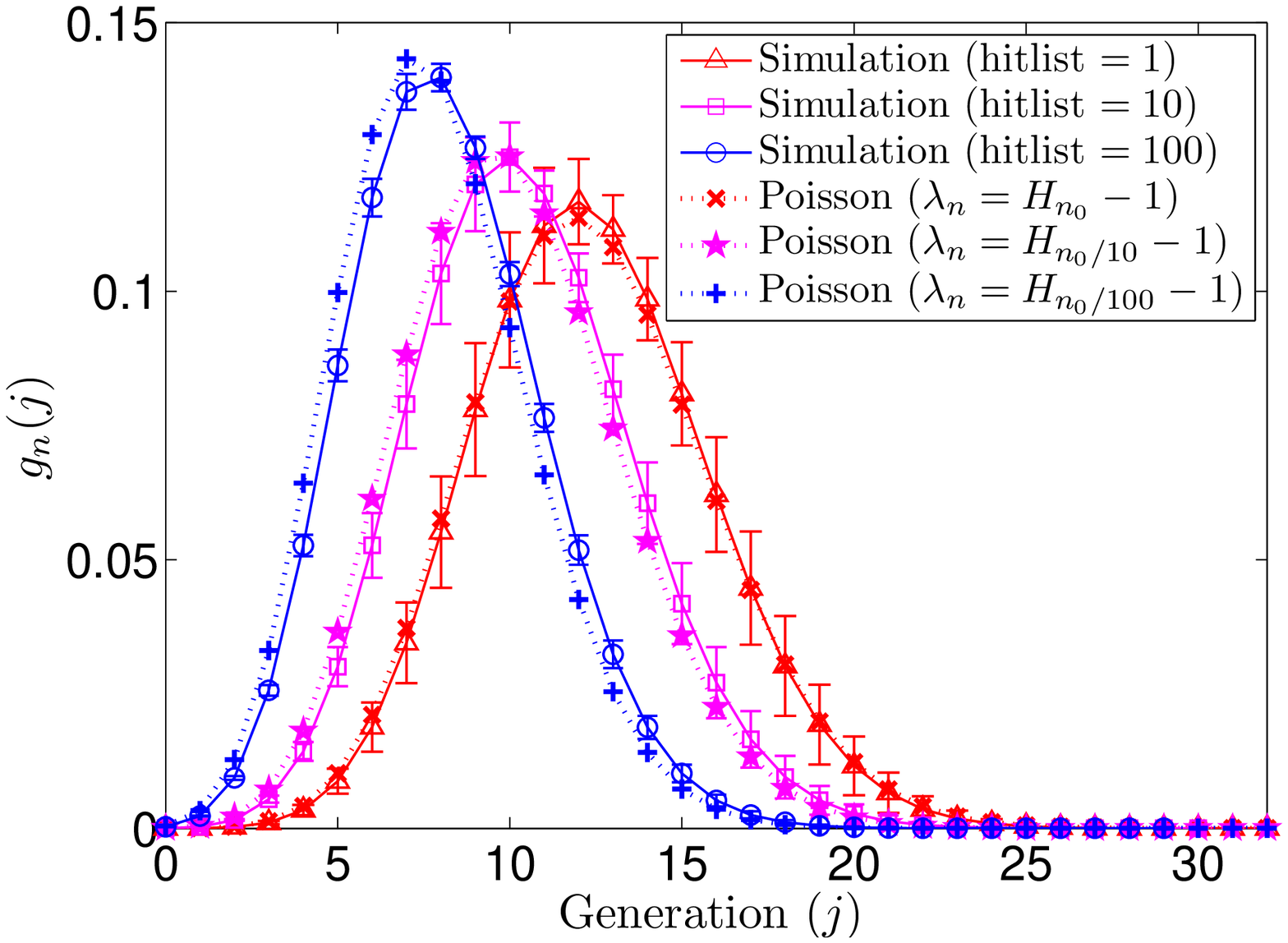}}
      }
       \caption{Effects of the scanning rate, the scanning rate standard deviation, and the hitlist size on
       the distributions of the number of children and the generation ($n_0=360000$).}
   \label{fig:simu_parameter}
\end{center}
\end{figure*}

Next, we extend our simulator to examine the effects of three
important parameters of worm propagation on the worm tree: the
scanning rate, the scanning rate standard deviation, and the hitlist
size. When a parameter is studied and varied, we set other
parameters to the parameters of the Code Red v2 worm as used in
Section \ref{sec:sim_ver}. The simulation results are obtained from
100 independent simulation runs and are shown in Fig.
\ref{fig:simu_parameter}.

Fig.s \ref{fig:simu_parameter}(a) and (b) show the effect of varying
the scanning rate $s$ (scans/min) from 158 to 558 on the
distributions of the number of children and the generation. Here,
the scanning rate is set to a fixed value for every infected host,
{\em i.e.,} the scanning rate standard deviation is 0. The figures
also plot the geometric distribution with parameter 0.5 and the
Poisson distribution with parameter $H_{n_0}-1$ for reference. It
can be seen that the scanning rate does not affect the worm tree
structure.
\begin{figure*}[tb]
\begin{center}
    \mbox{
         \hspace{-0.7cm}
      \subfigure[Number of children.]{\includegraphics[width=2.7in]{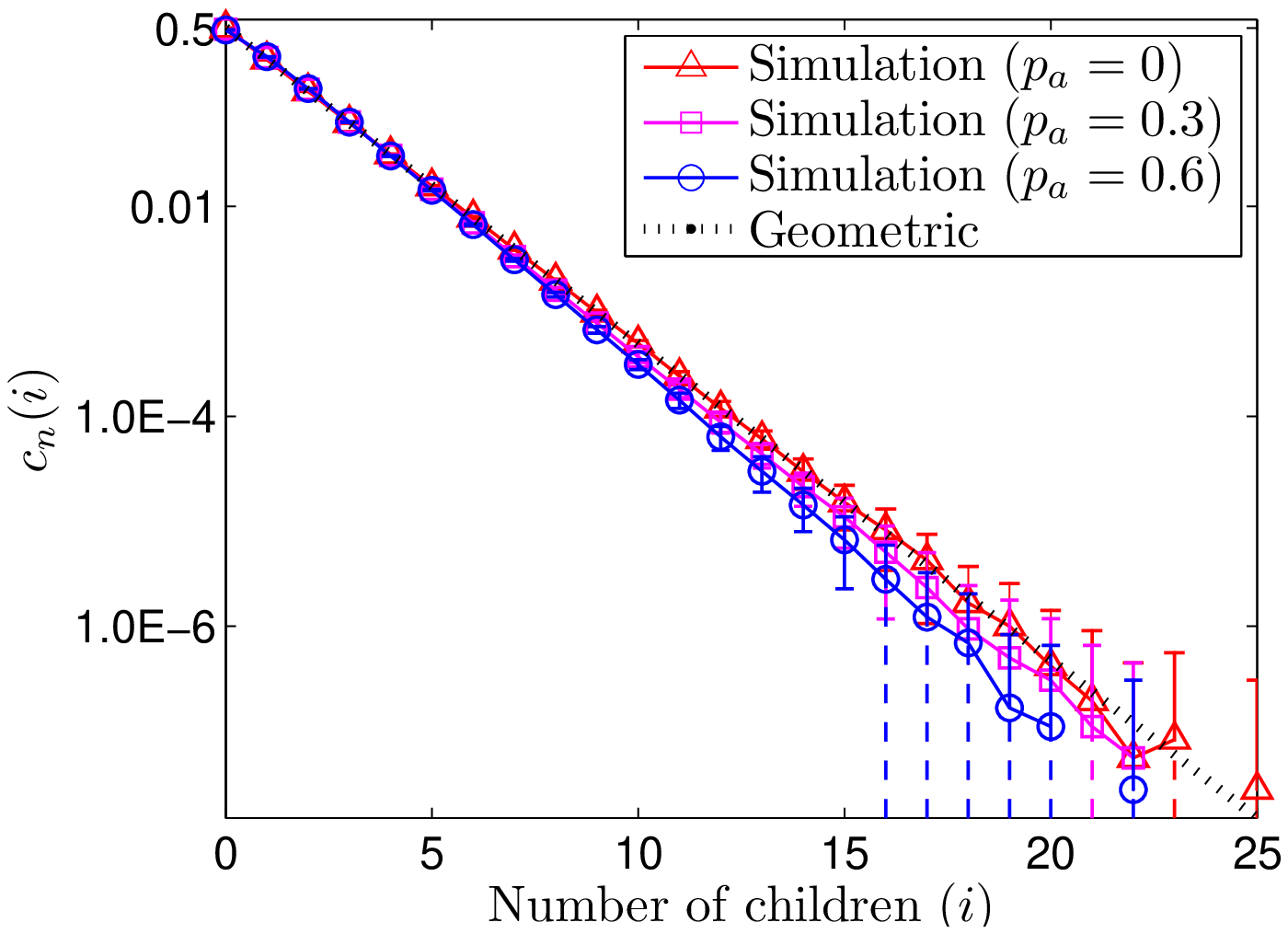}}
      \hspace{-0.7cm}
      \subfigure[Generation.]{\includegraphics[width=3.07in]{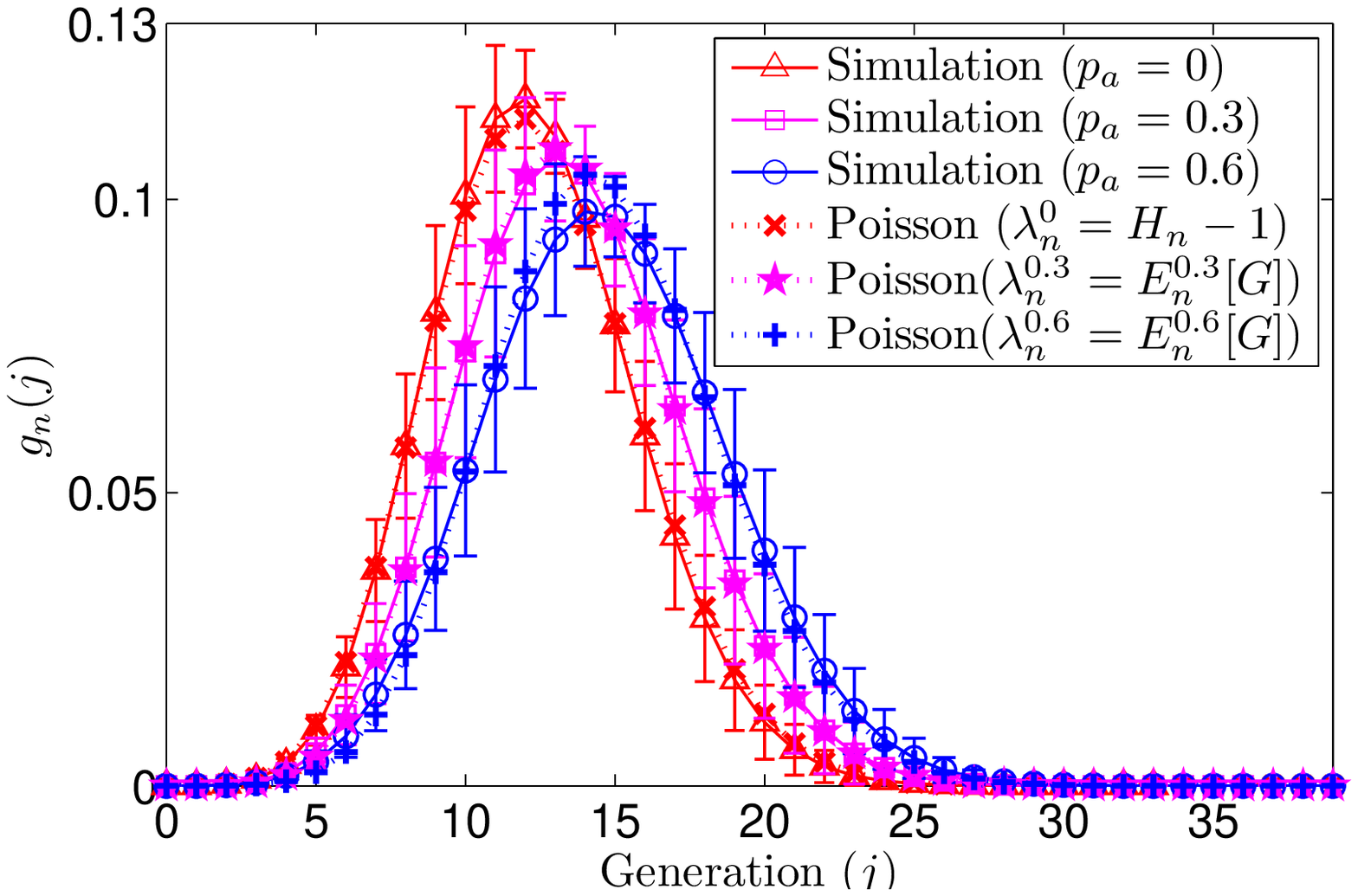}}
      \hspace{-0.9cm}
       \subfigure[Joint distribution ($p_a=0.6$).]{\includegraphics[width=2.37in]{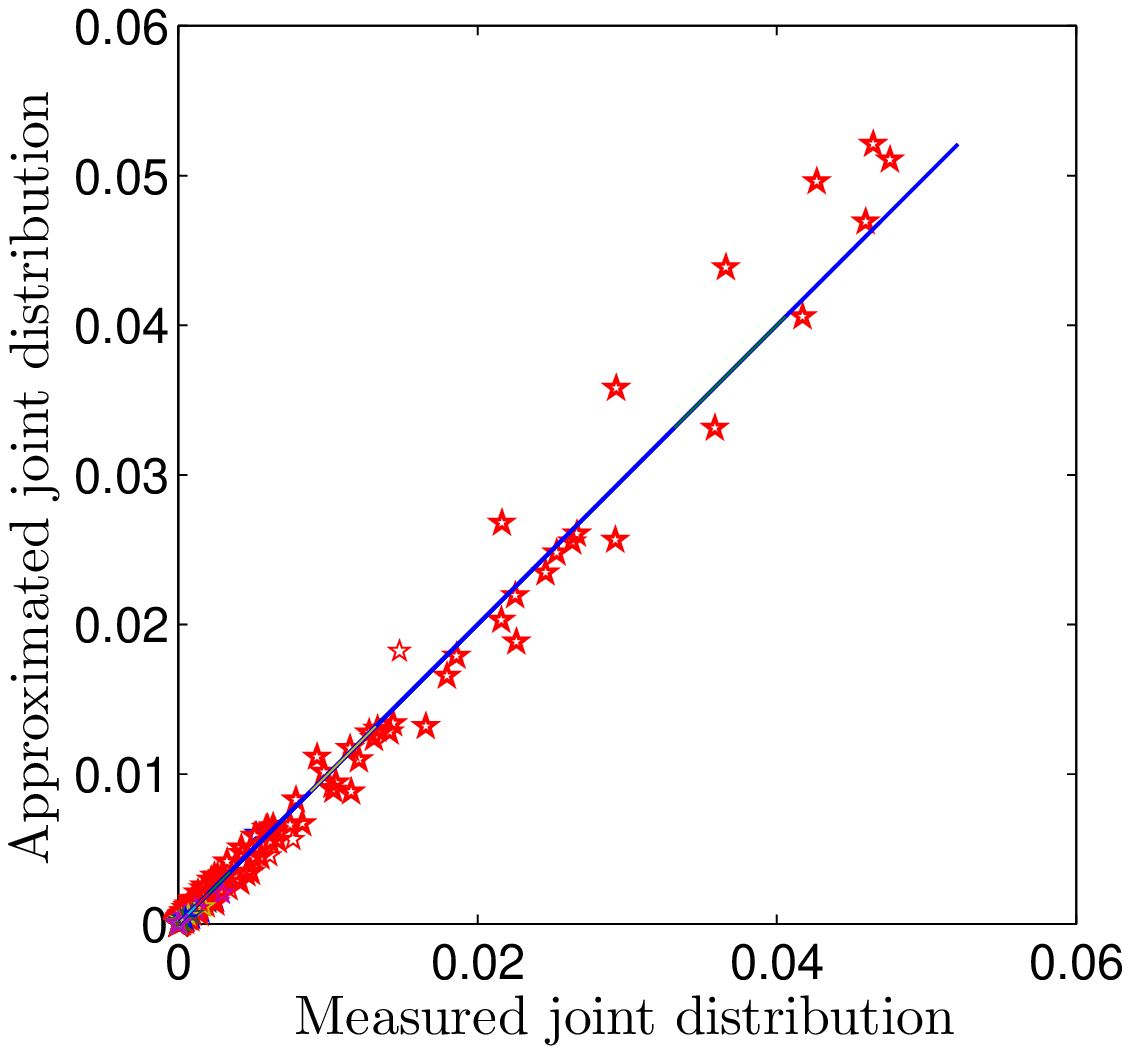}}
      }
          \caption{Simulating the infection structure of the localized-scanning worm ($n=360000$,
$s=358$ scans/min, $\sigma=0$, $hitlist=1$, and $l=8$).}
   \label{fig:simu_ls}
\end{center}
\end{figure*}

Fig.s \ref{fig:simu_parameter}(c) and (d) demonstrate the effect of
the variation of the scanning rates among different hosts ({\em
i.e.}, $\sigma$). In our simulation, a newly infected host is
assigned with a scanning rate (scans/min) from a normal distribution
$N(358,\sigma^2)$. The figures show the simulation results when
$\sigma=0$, $100$, and $200$. It can be seen that while the scanning
rate standard derivation $\sigma$ has no effect on the generation
distribution, it does affect the distribution of the number of
children. Specifically, when $\sigma$ increases, the tail of
$c_n(i)$ moves upward from the geometric distribution with parameter
0.5. This is because when $\sigma$ becomes larger, the variation of
the scanning rate among infected hosts is greater. That is, there
are more hosts with high scanning rates and also more hosts with low
scanning rates. As a result, those hosts with high scanning rates
tend to infect a large number of hosts, making the tail of $c_n(i)$
move upward. However, it is also observed that when $\sigma$ is not
very large (the case for real worms), the geometric distribution
with parameter 0.5 is still a good approximation.

In Fig.s \ref{fig:simu_parameter}(e) and (f), we show the effect of
the hitlist size on the worm tree. As pointed out in Section
\ref{sec:back}, when the hitlist size is greater than 1, the
underlying infection topology is a worm forest with the number of
trees equal to the hitlist size. Moreover, in a worm forest, it is
intuitive that each tree is a smaller version of the single worm
tree of hitlist size 1 and has fewer nodes.
Hence, it is not surprising to see that in Fig.
\ref{fig:simu_parameter}(f), the generation distribution moves
leftward when the hitlist size increases. However, the generation
distribution can still be well approximated by the Poisson
distribution with parameter $H_{n_h}-1$, where $n_h$ is the average
number of nodes in a tree. Moreover, since in each tree the
distribution of the number of children can be approximated by the
geometric distribution with parameter 0.5, in the worm forest
$c_n(i)$ still follows closely the same distribution.

\subsection{Localized Scanning} \label{sec:ls}

Finally, we extend our simulation study to the infection tree of
localized-scanning worms. Different from random scanning, localized
scanning preferentially searches for targets in the ``local" address
space \cite{Staniford}. As a result, when a new node is added to the
worm tree, it connects to one of the existing nodes that are in the
same ``local" address space with a higher probability. That is, the
growth model is no longer uniform attachment as studied in Section
\ref{sec:ana}. For simplicity, in this paper we only consider the
$/l$ localized scanning \cite{Chen07}:
\begin{itemize}
\item {\em Local scanning}:
$p_a (0 \leq p_a < 1)$ of the time, a ``local" address with the same
first $l$ ($0\le l \le 32$) bits as the attacking host is chosen as
the target.
\item {\em Global scanning}:
$1- p_a$ of the time, a random address is chosen.
\end{itemize}
Note that random scanning can be regarded as a special case of
localized scanning when $p_a = 0$. Moreover, if local scanning is
selected, it can be regarded as random scanning in a local $/l$
subnet. It has been shown that since the vulnerable-hosts
distribution is highly uneven, localized scanning can spread a worm
much faster than random scanning \cite{Rajab,Chen07infocom}.

We extend our simulator to imitate the spread of localized-scanning
worms. We extract the distribution of vulnerable hosts in $/l$
subnets from the dataset provided by DShield \cite{DShield,Paul:06}.
Specifically, we use the dataset in \cite{Paul:06} with port 80
(HTTP) that is exploited by the Code Red worm to generate the
vulnerable-host distribution. Moreover, we use similar
parameters as in Section \ref{sec:sim_ver} ({\em e.g.,} $n=360000$,
$s=358$ scans/min, $\sigma=0$, and $hitlist=1$) and set the subnet
level to 8 ({\em i.e.,} $l=8$). The results are obtained from 100
independent simulation runs and are shown in Fig. \ref{fig:simu_ls}.
For each run, patient zero is randomly chosen from vulnerable hosts.

Fig. \ref{fig:simu_ls}(a) compares the simulation results of the
distributions of the number of children ({\em i.e.,} $c_n(i)$) when
$p_a=0$, 0.3, and 0.6 with the geometric distribution with parameter
0.5. It is surprising that $c_n(i)$ of localized-scanning worms can
still be well approximated by the geometric distribution. That is,
the majority of nodes have few children, whereas a small portion of
compromised hosts infect a large number of hosts. An intuitive
explanation is given as follows. From Fig. \ref{fig:simu_rs}(a), it
can be seen that the total number of nodes has a minor effect on
$c_n(i)$. Hence, if in a /8 subnet the majority of vulnerable hosts
are infected through local scanning, it is expected that $c_n(i)$ of
these hosts still closely follows the geometric distribution since
the local scanning can be regarded as random scanning inside a /8
subnet. Therefore, both local infection and global infection lead
$c_n(i)$ towards the geometric distribution with parameter 0.5. On
the other hand, it can also be seen that when $p_a$ increases, the
tail of $c_n(i)$ moves slightly downward. This is because as $p_a$
increases, more vulnerable hosts are infected through local
scanning. Hence, it is more difficult for an infected host to find
targets after vulnerable hosts in this host's local subnet have been
exhausted. As a result, when $p_a$ increases, fewer nodes can have a
large number of children.

Fig. \ref{fig:simu_ls}(b) demonstrates that the generation
distribution of localized-scanning worms ({\em i.e.,} $g_n(j)$) can
be well approximated by the Poisson distribution for the cases of
$p_a=0$, 0.3, and 0.6. The Poisson parameter, however, depends not
only on $n$, but also on $p_a$. We further define $\lambda_n^{p_a} =
\mbox{E}_n^{p_a}[G]$ as the expectation of the generation for a
localized-scanning worm with parameter $p_a$. Here,
$\mbox{E}_n^{p_a}[G]$ can be easily estimated from the simulation
results of $g_n(j)$. Fig. \ref{fig:simu_ls}(c) further shows the
parity plot of the simulated joint distribution and the approximated
joint distribution from Equation (\ref{equ:joint_app}) when
$p_a=0.6$. Since most points are on or near the diagonal line, the
approximation is reasonable.
\begin{figure}[tb]
\begin{center}
{\includegraphics[width=3in]{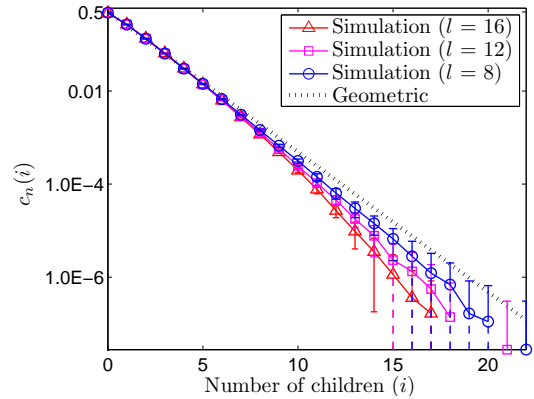}}
          \caption{Effect of the subnet level ($p_a=0.6$).}
          \label{fig:simu_ls_level}
\end{center}
\end{figure}

Moreover, Fig. \ref{fig:simu_ls_level} shows the effect of the
subnet level ({\em i.e.,} $l$) on the distribution of the number of
children ({\em i.e.,} $c_n(i)$). It can be seen that when $l$
increases, the tail of $c_n(i)$ moves downward. The reason is
similar to the argument used in Fig. \ref{fig:simu_ls}(a), {\em
i.e.,} as $l$ increases, fewer nodes can infect a large number of
children. However, the figure also demonstrates that the geometric
distribution with parameter 0.5 is still a good approximation to
$c_n(i)$, especially when the number of children is not large.

\section{Applications of Observations}
\label{sec:app}

Our observations on the topologies formed by worm infection have
important implications and applications for both defenders and
attackers. For example, we have found that the generation
distribution closely follows the Poisson distribution and the
average path length increases approximately logarithmically with the
number of nodes. On one hand, some schemes have been proposed to
trace worms back to their origins through the cooperation between
infected hosts \cite{Kumar05,Xie05}, and our work quantifies the
average path length that describes a lower bound of the number of
hosts required to cooperate. On the other hand, this average path
length reflects the delay or the effort for a botmaster to deliver a
command to all bots in a P2P-based botnet like Conficker C, and our
results show that the botnet is scalable and can efficiently forward
commands to a large number of bots. In this section, we focus on the
applications of the distribution of the number of children for both
defenders and attackers. Specifically, we study a simple and
efficient bot detection method in a Conficker C like P2P-based
botnet and consider a countermeasure by future botnets.

\subsection{Bot Detection}
\label{sec:bot_detection}

We consider a P2P-based botnet formed by worm scanning/infection.
That is, once a host infects another host, they become peers in the
resulting P2P-based botnet. When a defender captures an infected
host in a botnet, the defender can process the historic records
inside the host or monitor the traffic going into or out of the
host, and will potentially detect other infected hosts such as the
father and the children of this infected host. Then, our question is
that if a defender can only access a small portion of nodes in a
botnet, how many bots will be detected by the defender. Moreover,
inspired by the random removal and targeted removal methods used in
analyzing the robustness of a topology \cite{Barabasi_complex}, here
we study two bot detection strategies:
\begin{itemize}
\item Random detection: Access bots randomly.
\item Targeted detection: Access bots that have the largest number of children.
\end{itemize}

Analytically, we suppose that a defender can access a small ratio of
bots in a botnet. We assume that an accessed bot exposes itself, its
father, and its children to the defender. To simplify the analysis,
we also assume that the accessed bot ratio, $A$, is a power of 0.5
and all exposed nodes are different nodes. We then calculate the
average percentages of exposed bots by random detection and targeted
detection.

Since from Corollary \ref{cor:C} a randomly selected node has
approximately one child, the average percentage of bots that can be
exposed by random detection is then
\begin{equation}
\label{equ:random_detection}
  \textstyle   D_R = 3A.
\end{equation}

For targeted detection, since the nodes with the largest number of
children are chosen and the number of children follows
asymptotically a geometric distribution with parameter 0.5 as shown
in Corollary \ref{cor:geo},
\begin{equation}
  \textstyle  A = \sum_{i\ge d}{c_n(i)} =
  \sum_{i=d}^{\infty}{\left(\frac{1}{2}\right)^{i+1}} =
  \left(\frac{1}{2}\right)^d,
\end{equation}
where $d$ is the smallest number of children of accessed nodes. That
is, $d = -\log_2{A}$. Therefore, the average percentage of exposed
nodes by targeted detection is
\begin{equation}
\label{equ:targeted_detection}
   \textstyle  D_T = \sum_{i=d}^{\infty}{(2+i)\cdot
   c_n(i)}=(d+3)\left(\frac{1}{2} \right)^d = A(3-\log_2 A).
\end{equation}
Compared with random detection, targeted detection can expose
$(-A\log_2 A) \times n$ more nodes. For example, if
$A=\frac{1}{64}$, on average random detection can detect 4.69\% of
nodes, whereas targeted detection can expose 14.06\% of bots.
\begin{figure}[tb]
\begin{center}
{\includegraphics[width=3in]{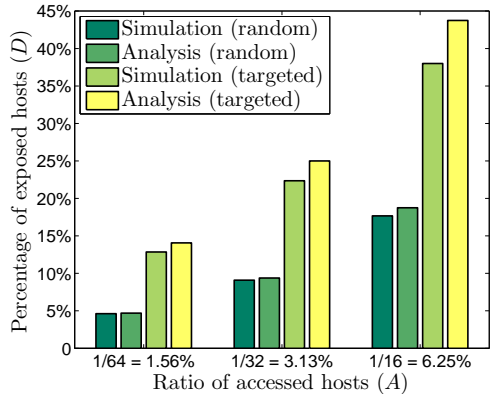}}
          \caption{Random and targeted detection.}
          \label{fig:removal}
\end{center}
\end{figure}

We then extend our simulation in Section \ref{sec:sim_ver} to study
the effectiveness of random and targeted detection strategies. Fig.
\ref{fig:removal} shows the simulation results over 100 independent
runs for both strategies, as well as the analytical results from
Equations (\ref{equ:random_detection}) and
(\ref{equ:targeted_detection}), when $A = \frac{1}{64}$,
$\frac{1}{32}$, and $\frac{1}{16}$. It can be seen that the
analytical results slightly overestimate the exposed host
percentage. This is because in our analysis we ignore the case that
two exposed nodes can be duplicate. Fig. \ref{fig:removal} also
demonstrates that targeted detection performs much better than
random detection. For example, in our simulation, when $A = 3.125\%$,
9.10\% of the bots are exposed under random detection, whereas
22.36\% of the bots are detected under targeted detection.
Therefore, when a small portion of bots are examined, the botnets
formed by worm infection are robust to random detection, but are
relatively vulnerable to targeted detection.

\subsection{A Countermeasure by Future Botnets}
\begin{figure*}[tb]
\begin{center}
    \mbox{
      \hspace{-0.6cm}
       \subfigure[Distribution of the number of children.]{\includegraphics[width=2.7in]{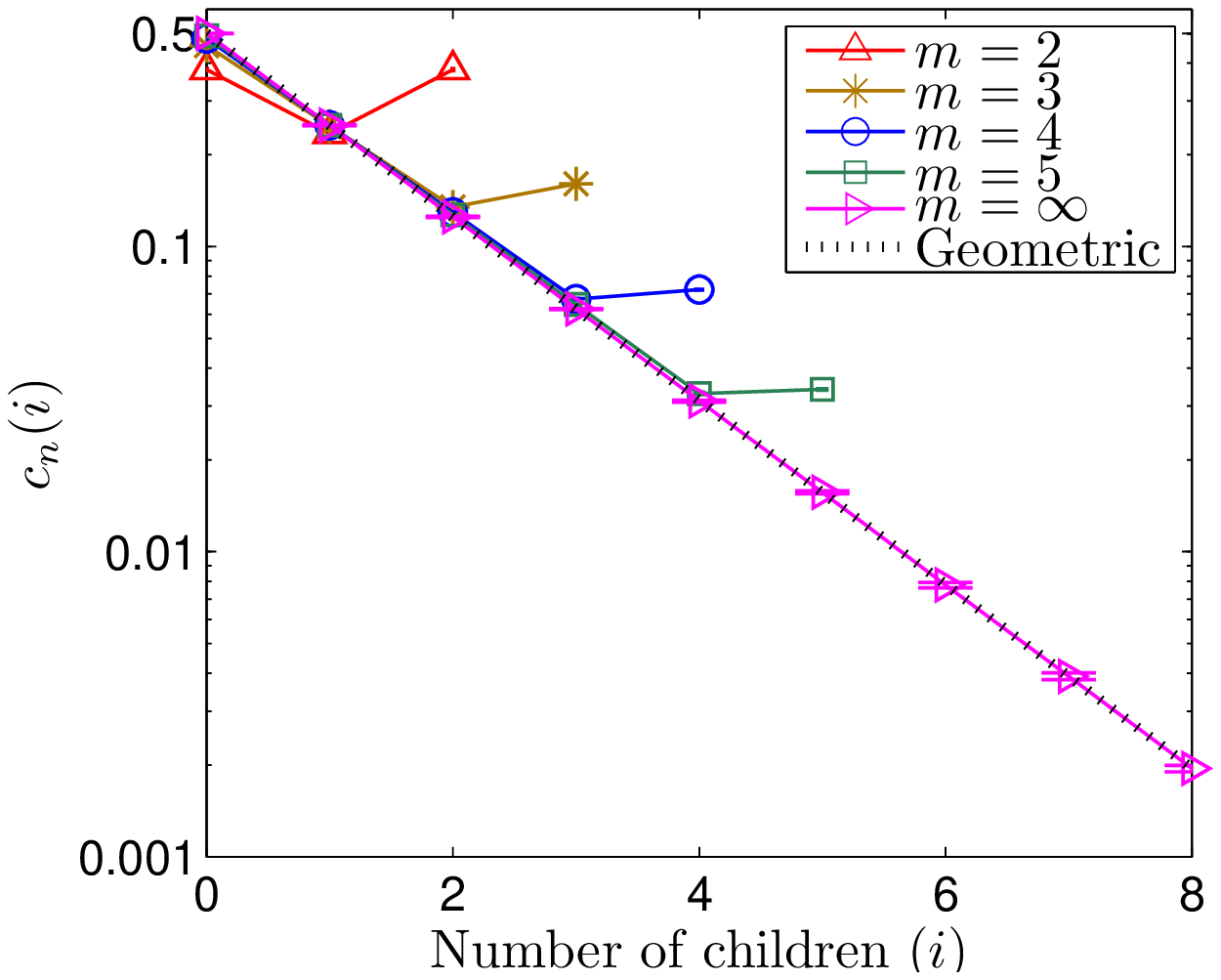}}
      \hspace{-1.2cm}
       \subfigure[Targeted detection.]{\includegraphics[width=2.7in]{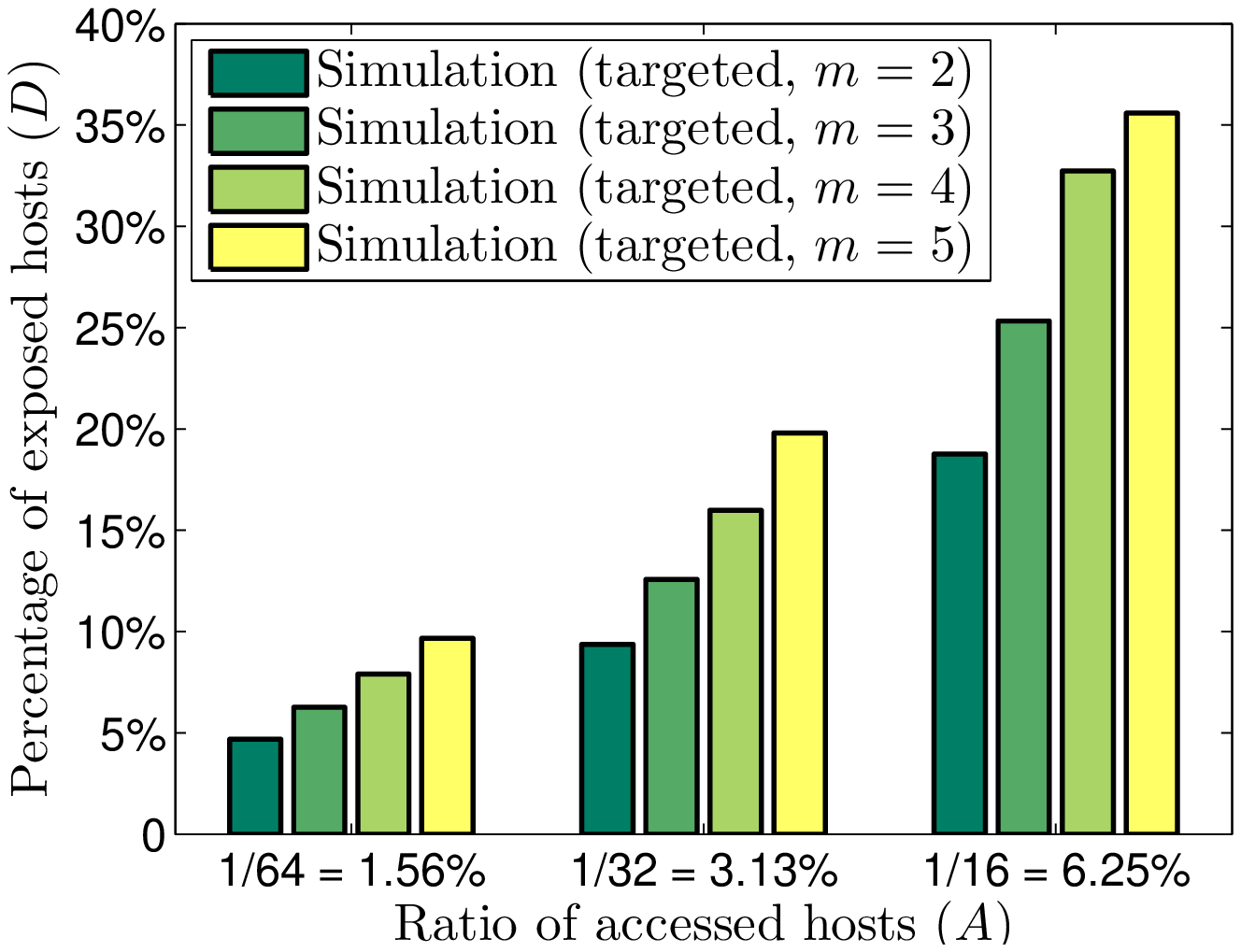}}
      \hspace{-0.7cm}
       \subfigure[Worm propagation speed.]{\includegraphics[width=2.7in]{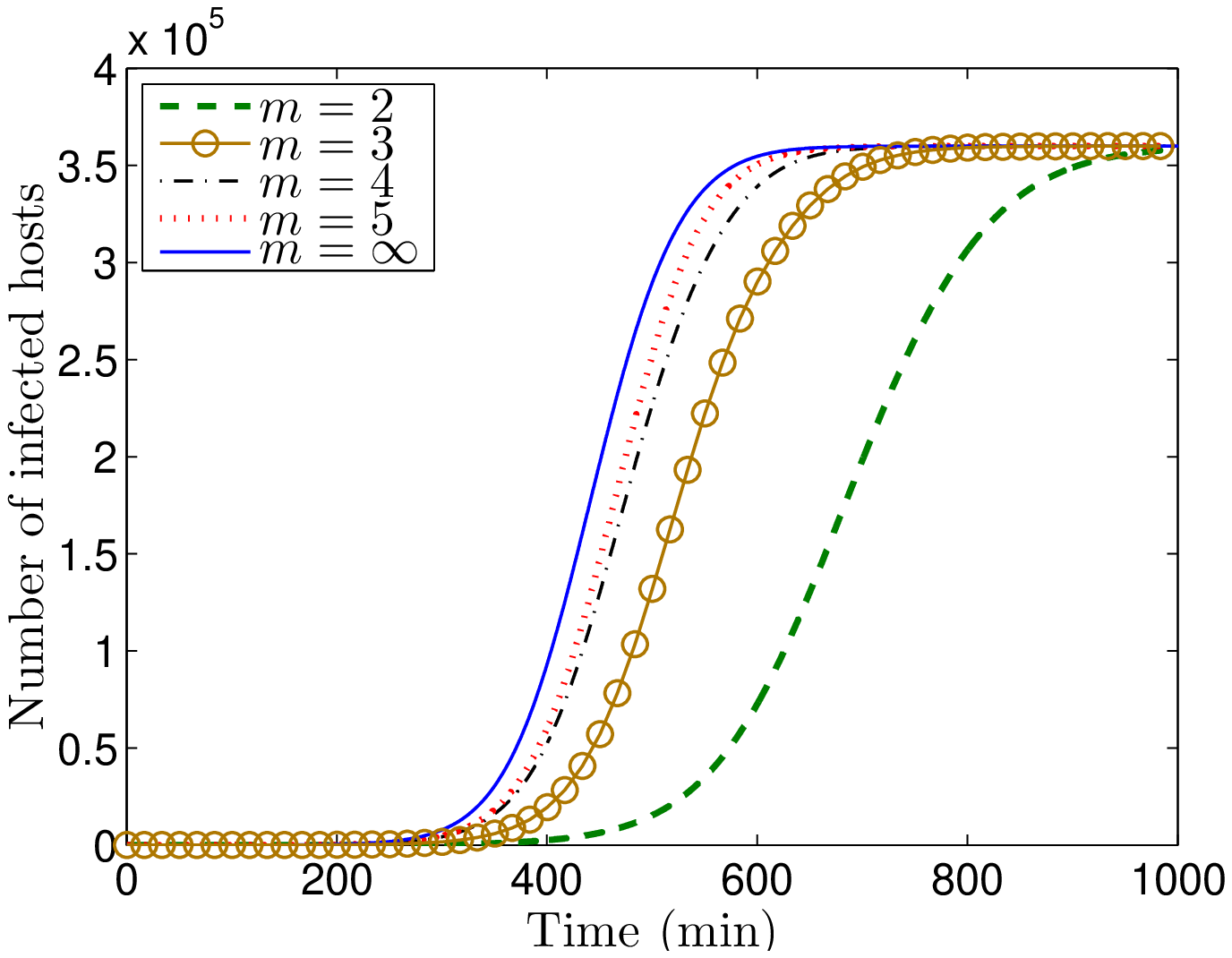}}
      }
          \caption{A worm countermeasure via limiting the maximum number of children.}
   \label{fig:simu_app}
\end{center}
\end{figure*}

To counteract the targeted detection method, an intuitive way for
botnets is to limit the maximum number of children for each node.
That is, set a small number $m$. Once an infected host has compromised $m$ other hosts, this
host stops scanning. In this way, there is no
node with a large number of children. Moreover, the infected hosts
can self-stop scanning, potentially reducing the worm traffic
\cite{Ma}.

To analyze the robustness of such botnets against targeted
detection, we extend Corollary \ref{cor:geo} to obtain an
approximated distribution of the number of children in a botnet with
the countermeasure:
\begin{equation}
\label{equ:countermeasure_dis}
\setlength{\extrarowheight}{0.1cm}
c_n(i)= \left\{\begin{array} {l l l}
\left(\frac{1}{2}\right)^{i+1},  & i = 0,1,2,\cdots,m-1 \\
\left(\frac{1}{2}\right)^m, & i = m.
\end{array} \right.
\end{equation}
The distribution is based on the observation that those nodes having
more than $m$ children in a botnet without the countermeasure can
now have only $m$ children. Hence, the expected percentage of
exposed nodes under targeted detection can be calculated:
\begin{equation}
\label{equ:countermeasure_det}
\setlength{\extrarowheight}{0.1cm}
D_T'= \left\{\begin{array} {l l l}
(m+2)\cdot A,  & A \le \left(\frac{1}{2} \right)^m \\
A(3-\log_2 A)- \left(\frac{1}{2}\right)^m, & A > \left(\frac{1}{2}
\right)^m.
\end{array} \right.
\end{equation}
Compared with $D_T$ in Equation (\ref{equ:targeted_detection}),
$D_T'$ is smaller. This means that under the countermeasure the number of exposed
nodes can be reduced significantly. For example, when $m=3$ and $A=\frac{1}{64}$,
$D_T=\frac{9}{64}$, and $D_T' = \frac{5}{64}$.

We then extend our simulation in Section \ref{sec:bot_detection} to
simulate the worm tree generated using the above countermeasure and
evaluate its performance against targeted detection. Fig.
\ref{fig:simu_app}(a) shows the distribution of the number of
children when $m=2$, 3, 4, and 5. It can be seen that except for $m=2$,
$c_n(i)$ is well approximated by Equation
(\ref{equ:countermeasure_dis}). For $m=2$, since an infected host
stops scanning when it has hit two vulnerable hosts, leaves in the
worm tree have more chances to recruit a child. Fig.
\ref{fig:simu_app}(b) demonstrates the expected percentage of
exposed nodes ({\em i.e.,} $D_T'$), when $A=\frac{1}{64}$,
$\frac{1}{32}$, and $\frac{1}{16}$, and $m=2$, 3, 4, and 5. It can
be seen that $D_T'$ follows approximately the analytical results in
Equation (\ref{equ:countermeasure_det}). Moreover, the expected
percentage of exposed nodes under the countermeasure is reduced
significantly. For example, when $A=\frac{1}{32}$, the percentage is
reduced from 22.36\% without the countermeasure to 19.80\%, 15.99\%,
12.58\%, and 9.38\% when $m=5$, 4, 3, and 2, respectively.

On the other hand, since not every infected host keeps scanning the
targets, the countermeasure can potentially slow down the speed of
worm infection. Thus, we also simulate the propagation speed of
worms that limit the maximum number of children and plot the results
in Fig. \ref{fig:simu_app}(c) for $m=2$, 3, 4, and 5, as well as the
original worm without the countermeasure. It can be seen that except for
$m=2$, the worm does not slow down much. But even when $m=2$, the
worm can infect most vulnerable hosts within 17 hours. Moreover,
Fig.s \ref{fig:simu_app}(b) and (c) demonstrate the tradeoff between
the efficiency of worm infection and the robustness of the formed
botnet topology. That is, a worm with the countermeasure spreads
slower, but the resulting botnet is more robust against targeted
detection.

\section{Related work}
\label{sec:related}

Since the Code Red worm in 2001, Internet worms have been an active
research topic. Many mathematical models have been developed to
characterize the spread of worms, estimate worm behaviors, and
contain worm propagation. Most models, however, have focused on the
{\em macro-level} behavior of worm infection. Specifically,
different analytical approaches have been applied to study the total
number of infected hosts over time
\cite{Staniford,Zou:scan,AAWP,Rohloff,Vojnovic,Dagon06,Zou05}. For
example, Staniford {\em et al.} used a simple differential equation
to estimate the global propagation speed of the Code Red v2 worm
\cite{Staniford}, whereas Rohloff {\em et al.} applied a stochastic
model to reflect the variation of the number of infected hosts at
the early stage of worm infection \cite{Rohloff}. The models of the
{\em micro-level} of worm infection, however, have been investigated
little. In this paper, we apply probabilistic modeling methods and
reveal some key micro-level information, such as the infection
ability of individual hosts and the underlying botnet topology
formed by worm infection.

Some efforts have been focused on studying the ``who infects whom"
information or the worm infection sequence
\cite{Kumar05,Rajab05,Xie05,Wang08}. Different from our work, the
prior work investigates the details of the random number generator
of worm propagation \cite{Kumar05} or infers the worm infection
sequence through the observations of network telescopes
\cite{Rajab05,Wang08}. Moreover, Sellke {\em et al.} applied a
branching process to study the effectiveness of a containment
strategy \cite{Sellke}. They assume that the total number of scans
of an infected host is bounded. As a result, the worm tree studied
in their work is fundamentally different from the one in our work.

Botnets have become the top threat to the Internet in recent years.
It has been shown that in current botnets, worm infection is still a
main tool for recruiting new bots or collecting network information,
and random scanning has been widely used \cite{Li09}. Moreover,
botnets are rapidly transiting from IRC systems to P2P systems. In
\cite{Ping}, Wang {\em et al.} gave a systematic study on P2P-based
botnets; whereas in \cite{Dagon_ACSAC'07_Taxonomy}, Dagon {\em et
al.} surveyed different P2P-based botnet topologies, such as random
graphs and power-law topologies. Several methods have been proposed
to construct P2P-based botnets through worm infection and
re-infection \cite{Vogt_NDSS'07_Army,Wang_HotBots'07_Hybrid}.

Modeling the topology generation process has been an active research
area. For example, Barab\'{a}si {\em et al.} developed the
well-known Barab\'{a}si-Albert (BA) model and used a mean-field
approach to characterize the growth of a topology with both
preferential attachment and uniform attachment
\cite{Barabasi_science,Barabasi_Physica_Mean}. Moreover, two exact
mathematical models have been studied for the BA model
\cite{Bollobas,Dorogovtsev}. From the theoretical aspect, our
proposed worm tree is similar to the random tree. For example,
Devroye used the records theory to derive the distribution of the
level of a random ordered tree in \cite{Luc}. Compared with these
theoretical efforts, our work studies a very different problem ({\em
i.e.,} botnets formed by worm infection) and uses a very different
approach ({\em i.e.,} probabilistic modeling).

\section{Conclusions}
\label{sec:conclusion}

In this paper, we attempt to capture the key characteristics of the
Internet worm infection family tree and apply them to bot detection.
We have shown analytically and empirically that for the infection
tree formed by a wide class of worms, the number of children
asymptotically has a geometric distribution with parameter 0.5; and
the generation closely follows a Poisson distribution with parameter
$\mbox{E}_n[G]$ ({\em i.e.,} $H_n-1$). As a result, on average half
of infected hosts never compromise any target, over 98\% of nodes
have no more than five children, and a small portion of hosts have a
large number of children. Moreover, the average path length of the
worm tree increases approximately logarithmically with the number of
nodes. We have also demonstrated empirically that similar
observations can be found in localized-scanning worms. We have then
applied the observations to bot detection and found that targeted
detection is an efficient way to expose bots in a botnet. However,
we have also pointed out that a simple countermeasure by future
botnets can weaken the performance of targeted detection, without
greatly slowing down the speed of worm infection.

As part of our ongoing work, we plan to study in more depth
efficient methods against future botnets and relax our assumptions
to include more worm dynamics. For example, we are studying the
effect of user defenses on the worm tree \cite{Wang:globecom10}.

\section*{APPENDIX 1: Proof of Corollary 1}  % use *-form to suppress numbering
%\begin{small}
%\renewcommand{\theequation}{A-\arabic{equation}}
%\setcounter{equation}{0}  % reset counter

\label{app:cor1}

We apply z-transform to derive the expectation and the variance of
the number of children. First, note that Corollary \ref{cor:C} holds
for $n=1$ and $2$. Next, when $n\ge 3$, we define z-transform
\begin{equation}
\label{equ:zx}
  \textstyle   X_n(z)=\sum_{i=0}^{n-1}{c_n(i)z^{-i}}.
\end{equation}
Setting $c_{n-1}(-1)=1$, we can transform Theorem \ref{thm:pc} to
\begin{equation}
\label{equ:rc_z}
  \textstyle   c_n(i) = \frac{n-2}{n}c_{n-1}(i)+\frac{1}{n}c_{n-1}(i-1),\ 0\le
    i \le n-1,
\end{equation}
when $n \ge 3$. Then, putting Equation (\ref{equ:rc_z}) into
Equation (\ref{equ:zx}), we can obtain the difference equation of
z-transform
\begin{equation}
  \textstyle   X_n(z) = \left(\frac{1}{n}z^{-1}+\frac{n-2}{n}
    \right)X_{n-1}(z)+\frac{1}{n}.
\end{equation}
Note that $\mbox{E}_n[C]=-\frac{dX_n(z)}{dz}\mid_{z=1}$ and
$X_{n-1}(1)=1$, which leads to
\begin{equation}
  \textstyle   \mbox{E}_n[C] = \frac{n-1}{n}\mbox{E}_{n-1}[C]+\frac{1}{n}.
\end{equation}
Since $\mbox{E}_2[C]=\frac{1}{2}$, we can show by induction that
\begin{equation}
  \textstyle   \mbox{E}_n[C]=\frac{n-1}{n}.
\end{equation}

Moreover, $\mbox{E}_n[C^2]=\frac{d}{dz}\left[z\frac{dX_n(z)}{dz}
\right]\mid_{z=1}$ yields
\begin{eqnarray}
  \mbox{E}_n[C^2] & = &   \textstyle
    \frac{n-1}{n}\mbox{E}_{n-1}[C^2]+\frac{2}{n}\mbox{E}_{n-1}[C]+\frac{1}{n}
    \\
   & = & \textstyle  \frac{n-1}{n}\mbox{E}_{n-1}[C^2]+\frac{3n-5}{n^2}.
\end{eqnarray}
Thus, we can use $\mbox{E}_2[C^2]=\frac{1}{2}$ to prove by induction
that
\begin{equation}
  \textstyle   \mbox{E}_n[C^2] = 2+\frac{(n-1)(n-2)}{n^2}-\frac{2H_n}{n},
\end{equation}
where $H_n=\sum_{i=1}^n{\frac{1}{i}}$ is the $n$-th harmonic number
\cite{Harmonic}. Therefore,
\begin{eqnarray}
        \mbox{Var}_n[C] & = &  \textstyle \mbox{E}_n[C^2]-\mbox{E}_n^2[C] \\
                        & = &  \textstyle  2-\frac{n-1}{n^2}-\frac{2H_n}{n}.
\end{eqnarray}

\section*{APPENDIX 2: Proof of Corollary 2}  % use *-form to suppress numbering

It is already known that $c(0)=\frac{1}{2}$. When $i\geq 1$, this
corollary follows readily from Equation (\ref{equ:rc}). Since
$n\rightarrow \infty$, $ c_{n-1}(i) = c_{n}(i) = c(i)$, which yields
\begin{equation}
    c(i) = \textstyle \frac{n-2}{n}c(i) + \frac{1}{n}c(i-1).
\end{equation}
That is,
\begin{equation}
    c(i) =  \textstyle  \frac{1}{2}c(i-1),\ \  i\geq 1.
\end{equation}
Hence, from $c(0)=\frac{1}{2}$, we can recursively obtain
\begin{equation}
    c(i) = \textstyle  \big(\frac{1}{2}\big)^{i+1},\ \  i \geq 0.
\end{equation}

\section*{APPENDIX 3: Proof of Corollary 3}

Similar to the proof of Corollary \ref{cor:C}, we apply z-transform
to derive the expectation and the variance of the generation. First,
note that Corollary \ref{cor:G} holds for $n=1$ and $2$. Next, when
$n\ge 3$, we define z-transform
\begin{equation}
\label{equ:zxg}
 \textstyle    Y_n(z)=\sum_{j=0}^{n-1}{g_n(j)z^{-j}}.
\end{equation}
Putting Equation (\ref{equ:rg}) into Equation (\ref{equ:zxg}), we
can obtain the difference equation of z-transform
\begin{equation}
\textstyle    Y_n(z) = \left(\frac{1}{n}z^{-1}+\frac{n-1}{n}
    \right)Y_{n-1}(z).
\end{equation}
Note that $\mbox{E}_n[G]=-\frac{dY_n(z)}{dz}\mid_{z=1}$ and
$Y_{n-1}(1)=1$, which leads to
\begin{equation}
    \textstyle \mbox{E}_n[G] =\mbox{E}_{n-1}[G]+\frac{1}{n}.
    \label{eqn:pgm1}
\end{equation}
Since $\mbox{E}_2[G]=\frac{1}{2}$, we can show by induction that
\begin{equation}
    \mbox{E}_n[G]=H_n -1.
\end{equation}

Moreover, $\mbox{E}_n[G^2]=\frac{d}{dz}\left[z\frac{dY_n(z)}{dz}
\right]\mid_{z=1}$ yields
\begin{equation}
   \textstyle  \mbox{E}_n[G^2]  =
\mbox{E}_{n-1}[G^2]+\frac{2}{n}\mbox{E}_{n-1}[G]+\frac{1}{n}.
\label{eqn:pgm2}
\end{equation}
Therefore, combining Equations (\ref{eqn:pgm1}) and (\ref{eqn:pgm2})
gives
\begin{eqnarray}
\mbox{Var}_n[G]& = &\mbox{E}_n[G^2] - \mbox{E}^2_n[G]  \nonumber \\
&=& \mbox{E}_{n-1}[G^2]   +\textstyle \frac{1}{n}( 2\mbox{E}_{n-1}[G]+1)  \nonumber \\
&& \textstyle  -(\mbox{E}_{n-1}[G] + \frac{1}{n})^2\nonumber \\
&=& \textstyle  \mbox{Var}_{n-1}[G] + \frac{1}{n} - \frac{1}{n^2}.
\end{eqnarray}
Thus, we can use $\mbox{Var}_2[G]=\frac{1}{4}$ to prove by induction
that
\begin{equation}
\mbox{Var}_n[G] = H_{n}-H_{n,2},
\end{equation}
where $H_n=\sum_{i=1}^n{\frac{1}{i}}$ and $H_{n,2}
=\sum_{i=1}^n{\frac{1}{i^2}}$.

\section*{APPENDIX 4: Proof of Corollary 4}

We prove this corollary by applying z-transform. If a random
variable $X$ follows a Poisson distribution with parameter
$\lambda$,
\begin{equation}
    \mbox{Pr}(X=k) = \frac{\lambda^k}{k!}e^{-\lambda}, \ \
    k=0,1,2,\cdots.
\end{equation}
Using z-transform, we have
\begin{equation}
    X(z) = \sum_{k=0}^{\infty}\mbox{Pr}(X=k)z^{-k}=e^{\lambda
    \left(z^{-1}-1\right)}.\label{eqn:psz}
\end{equation}
Meanwhile, using Equation (\ref{equ:rg}) in Theorem \ref{thm:pg}, we
find the z-transform of $g_n(j)$
\begin{equation}
\textstyle
Y_n(z)=\sum_{j=0}^{n-1}{g_n(j)z^{-j}}=\left(1+\frac{z^{-1}-1}{n}
    \right)Y_{n-1}(z).
\end{equation}
Note that when $x \rightarrow 0$, $e^x\approx 1+x$. Thus, when $n$
is very large, $1+\frac{z^{-1}-1}{n}\approx
\mbox{exp}((z^{-1}-1)/n)$. That is,
\begin{equation}
    Y_n(z) \approx e^{\frac{z^{-1}-1}{n}}Y_{n-1}(z).
\end{equation}
Using $Y_1(z)=1$, we can recursively obtain
\begin{equation}
    Y_n(z)\approx e^{(z^{-1}-1)\sum_{i=2}^n{\frac{1}{i}}} =
    e^{(H_n-1)(z^{-1}-1)}. \label{eqn:gz}
\end{equation}
Therefore, by comparing Equations (\ref{eqn:psz}) and
(\ref{eqn:gz}), $g_n(j)$ can be approximated by the Poisson
distribution with parameter $H_n-1$ as in Equation (\ref{equ:rg_P}).

\bibliography{./TDSC}
%
%\begin{IEEEbiography}
%%[{\includegraphics[width=1in,height=1.25in,clip,keepaspectratio]{./RandyPausch}}]
%{Qian Wang} received his B.E. degree from Dalian University of
%Technology, China in 2005 and his M.S. degree from Florida
%International University in 2008. Currently he is working towards
%the Ph.D. degree in the Department of Electrical and Computer
%Engineering at Florida International University. His research
%interests include network security and statistical modeling.
%\end{IEEEbiography}
%
%\begin{IEEEbiography}
%%[{\includegraphics[width=1in,height=1.25in,clip,keepaspectratio]{./RandyPausch}}]
%{Zesheng Chen} is an associate faculty member with the Department of
%Engineering, Indiana University - Purdue University Fort Wayne.
%He received his M.S. and Ph.D. degrees from the School
%of Electrical and Computer Engineering at the Georgia Institute of
%Technology in 2005 and 2007, respectively. His research interests
%include network security and the performance evaluation of computer
%networks.
%\end{IEEEbiography}
%
%\begin{IEEEbiography}
%%[{\includegraphics[width=1in,height=1.25in,clip,keepaspectratio]{./RandyPausch}}]
%{Chao Chen} is an assistant professor with the Department of
%Engineering, Indiana University - Purdue University Fort Wayne. She
%received her M.S. and Ph.D. degrees from Georgia Institute of
%Technology in 2003 and 2005, respectively. Her current research
%interests include routing in mobile ad hoc networks and space-based
%communication networks, modeling and performance evaluation of
%wireless opportunistic networks, and network security.
%\end{IEEEbiography}

\end{document}